\documentclass[iop,twocolappendix]{emulateapj}
\usepackage{textcase}
\usepackage{amssymb,amsmath}
\usepackage{graphicx}
\usepackage{natbib}
\usepackage{float}
\usepackage{multirow}
\newcommand{\penn}{1}
\newcommand{\jpl}{2}

\shorttitle{[CII] Line Intensity Mapping During $0.5 < \MakeTextLowercase{z} < 1.5$}
\shortauthors{Uzgil, Aguirre, Bradford, \& Lidz}

\begin{document}

\title{Measuring galaxy clustering and the evolution of [CII] mean intensity with Far-IR line intensity mapping during $0.5 < \MakeTextLowercase{z} < 1.5$}

\author{B.~D.~Uzgil\altaffilmark{\penn,\jpl}}

\author{J.~E.~Aguirre\altaffilmark{\penn}}

\author{C.~M.~Bradford\altaffilmark{\jpl}}

\author{A.~Lidz\altaffilmark{\penn}}

\email{badeu@sas.upenn.edu}

\altaffiltext{\penn}{University of Pennsylvania, Philadelphia, PA 19104}

\altaffiltext{\jpl}{Jet Propulsion Laboratory}

\begin{abstract}

Infrared fine-structure emission lines from trace metals are powerful diagnostics of the interstellar medium in galaxies. We explore the possibility of studying the redshifted far-IR fine-structure line emission using the three-dimensional (3-D) power spectra obtained with an imaging spectrometer.  The intensity mapping approach measures the spatio-spectral fluctuations due to line emission from all galaxies, including those below the individual detection threshold. The technique provides 3-D measurements of galaxy clustering and moments of the galaxy luminosity function.  Furthermore, the linear portion of the power spectrum can be used to measure the total line emission intensity including all sources through cosmic time with redshift information naturally encoded.   Total line emission, when compared to the total star formation activity and/or other line intensities reveals evolution of the interstellar conditions of galaxies in aggregate.  As a case study, we consider measurement of [CII] autocorrelation in the $0.5 < z < 1.5$ epoch, where interloper lines are minimized, using far-IR/submm balloon-borne and future space-borne instruments with moderate and high sensitivity, respectively.  In this context, we compare the intensity mapping approach to blind galaxy surveys based on individual detections.  We find that intensity mapping is nearly always the best way to obtain the total line emission because blind, wide-field galaxy surveys lack sufficient depth and deep pencil beams do not observe enough galaxies in the requisite luminosity and redshift bins.  Also, intensity mapping is often the most efficient way to measure the power spectrum shape, depending on the details of the luminosity function and the telescope aperture.

\end{abstract}

\keywords{far-infrared spectroscopy; galaxy redshift surveys; large-scale structure}

\section{Introduction}

Charting the history throughout cosmic time of star formation, black hole growth, and the properties of the galaxies that host these activities is at the root of many astronomical measurements currently underway. A fundamental limitation of most galaxy surveys---both photometric and spectroscopic---is that they are flux-limited, translating to a threshold luminosity below which galaxies are not included in the observations. This incompleteness is particularly true in the far-infrared/submillimeter wavelengths, which seem to have dominated the historical energy output of galaxies \citep{planckXXX}. With the exception of ALMA, which is not well-suited to large surveys, telescopes remain sensitivity-challenged in this regime.

Intensity mapping by its nature probes all sources of emission, whether point-like or diffuse, luminous or faint. We focus here on three-dimensional (3-D) line intensity mapping, also known as tomographic mapping, using the spatial and spectral dimensions. A 3-D intensity mapping survey targeting a spectral line at a range of frequencies naturally produces a data cube in which redshift, thus line-of-sight distance is automatically encoded. The 3-D fluctuations in line emission are then studied in Fourier space with the power spectrum. This approach expands upon recent works that utilize the fluctuations in emission (rather than individually detected galaxies with luminosities down to a survey's flux limit) to study the properties of dusty, star-forming galaxies (DSFGs) with continuum data. These studies, using \emph{P(D)} \citep{glenn10, bethermin11} or a 2-D power spectrum \citep{viero13, planckXXX} analysis, have already shed light on some aspects (such as galaxy number counts, spatial clustering, and cosmic evolution of IR luminosity density) of the bulk of these systems during the peak of cosmic star formation, but they are limited by source confusion or uncertainties associated with the lack of redshift information. Redshift ambiguities can be removed to some extent with galaxy-by-galaxy observations with the interferometers ALMA or NOEMA, or with an instrument like X-Spec, a proposed multi-object spectrometer for CCAT. However, the interferometer surveys will be expensive and will cover very little sky, and the CCAT surveys, though faster, will not reach the faintest galaxies in the luminosity function \citep{bradford09ccat}. Power spectrum treatment of the 3-D datasets naturally combines the redshift precision of spectral measurements, while including all sources of emission, and can be carried out with an instrument that does not require exquisite point-source sensitivity. 

Atomic \citep{gongcii,visbal11,suginohara99} and molecular \citep{lidz11,gong11co} transitions---such as the 21-cm spin flip transition from H$^{\mathrm{o}}$, CO rotational lines, and [CII]158$\mu$m---have been investigated as candidates for intensity mapping experiments during the Epoch of Reionization (EoR) and afterward \citep[][for CO lines and Ly$\alpha$]{pullen13lya,pullen13co, breysse14}. Of these, the neutral hydrogen case is undoubtedly the most developed in terms of its standing in the literature (cf. \citet{morales10rev} for a review) and in the experimental arena (e.g., PAPER \citep{parsons14}, MWA \citep{tingay13}) because intensity mapping is the only means of studying the intergalactic HI light. [CII] later emerged as an EoR intensity mapping candidate since it both offers a way to probe the clustering of sources from the faint-end of the luminosity function, and provides an opportunity for cross-correlation with the HI datasets \citep{gongcii}.

In addition to tracing large-scale structure, [CII] also contains astrophysical information about the conditions in star-forming galaxies. With an ionization potential of 11.6 eV, it arises in both ionized and neutral atomic gas.  Empirically, it is an important coolant, often the brightest single line in the spectrum of a star-forming galaxy, emitting as much as 0.5--1\% of the total far-IR luminosity \citep{malhotra97,luhman98,stacey10,graciacarpio11}.   The ratio of the [CII] luminosity to the total bolometric luminosity can be used as a diagnostic tool that provides: (1) a measure of the star-formation activity, (2) a measure of the spatial extent (or ``mode") of star formation, and (3) an AGN/starburst discriminant \citep{hailey-dunsheath10,stacey10,graciacarpio11,sargsyan12,diaz-santos13}.

The broader suite of far-IR lines probes all phases of the interstellar medium, and the negligible optical depth of galaxies at far-IR wavelengths ensures that even the most heavily embedded regions where stars form and black holes grow are revealed. For the atomic and ionized medium, the key far-infrared emission lines are those of C, N, \& O (e.g., [OI]63$\mu$m, 146$\mu$m, [CII]158$\mu$m, [OIII]52$\mu$m, 88$\mu$m, [NIII]57$\mu$m, and [NII]122$\mu$m, 205$\mu$m). The emitting species cover more than an order of magnitude in ionization potential and they strongly constrain the density and temperature of the ionized and neutral gas, and the strength and hardness of the interstellar radiation field.  These physical parameters then reveal the relative importance of the black hole vs. the hot young stars to the overall energy budget, and constrain the stellar effective temperatures \cite[e.g.]{rubin85,dale04,colbertM82,malhotra01,ferkinhoff11,lebouteiller12}. The suite of carbon, oxygen and nitrogen transitions also measure abundances \citep{garnett04, lester87,nagao11}.

Line intensity mapping experiments targeting the fine-structure metal lines at post-Reionization redshifts can offer a proof of principle of the approach, similar to measurements of the HI autocorrelation power spectrum at $z\sim0.8$ \citep{chang10, switzer13}, and provide a complete census of galaxies during an important phase in the star formation history of the Universe. While the redshifted far-IR lines are not accessible from the ground in this redshift range, a balloon- or space-borne intensity mapping experiment with broad wavelength coverage can in principle measure the mean intensities of these lines through cosmic time, thereby charting the evolution of the star-formation conditions in galaxies in an absolute, aggregate sense.   Here we consider a first step in this direction: a measurement of [CII] autocorrelation in multiple bins through the $0.5 < z < 1.5$ epoch.   [CII] and far-IR lines in general ought to be particularly well-suited to this time frame, as $z\sim1.5$ is believed to be the peak in the dust attenuation in galaxies, when roughly 80\% of the cosmic star formation rate density is obscured and captured only in the infrared emission of re-processed starlight by dust grains \citep{burgarella13}.   From a practical standpoint, [CII] in this epoch is relatively free of interloper lines, as will be shown.

The organization of this paper is as follows. We have estimated the mean intensity for a suite of fine-structure IR emission lines, including the [CII] line, based on empirical IR luminosity functions and line-to-IR luminosity correlations, and present these results in the context of a power spectrum model in Section 2. In Section 3, we envision suitable platforms for conducting the [CII] intensity mapping experiment and discuss the feasibility of detecting the [CII] power spectra in terms of the signal-to-noise ratio (SNR). From the predicted power spectra, we provide estimates for accuracy in measuring the mean [CII] intensity as a function of redshift. To better assess the value of intensity mapping studies in the case of [CII] at moderate redshifts, and of intensity mapping experiments in general, we compare in Section 4 the performance of the intensity mapping approach against spectroscopic galaxy surveys that rely on individual detections of sources to measure the total emission and power spectrum. In particular, we examine the effects of variations in luminosity function shape, aperture diameter (and, consequently, voxel size), and experimental noise on the ability of each observational method to measure the power spectrum and provide a complete view of the galaxy population. 

\section{Predictions for Far-IR Line Power Spectra}

\subsection{Relationship Between Galaxy Populations and Fluctuation Power}

The complete autocorrelation power spectrum of a given far-IR fine-structure line $i$ as a function of wavenumber $k$, $P_{i,i}(k,z)$, can be separated into power from the clustering of galaxies, $P_{i,i}^{clust}(k,z)$ and a Poisson term arising from their discrete nature, $P_{i,i}^{shot}(z)$. We compute the full nonlinear matter power spectrum, $P_{\delta,\delta}(k, z)$, using the publicly available code HALOFIT+ (http://camb.info), which has been the standard tool for predicting matter power spectra upon its success in fitting state-of-the-art dark matter simulations over a decade ago \citep{halofit}.  The clustering component of the line power spectrum is then written as \citep{visbal10}
\begin{equation}
 P_{i,i}^{clust}(k, z) = \bar{S}_{i}^2(z) \bar{b_i}^2(z) P_{\delta\delta}(k, z).
 \label{eq:pclust}
 \end{equation}
Here we implicitly assume that the fluctuations in line emission trace the matter power spectrum with some linear bias, $\bar{b_i}(z)$, but note that we use the full nonlinear matter power spectrum. This should be an adequate approximation for our study, since the Poisson term (see Equation~\ref{eq:pshot}) will dominate on small-scales where the non-linearities become significant. For our target redshift range and likely [CII] emitters, $\bar{b}_i$ is reasonably well-constrained to be between 2 and 3 \citep{Cooray10, Jullo12}, so we have assumed a single bias at each redshift, although a more sophisticated model would allow for variation of the source bias with the host halo mass (and thus luminosity). It should be straightforward to be rescale the results for other assumptions about bias. 

The mean line intensity, $\bar{S}_{i}(z)$, in units of Jy sr$^{-1}$, can be calculated as
\begin{equation}
\bar{S}_{i}(z) = \int{\mathrm{d}n_{i} \frac{L_{i}}{4\pi D_L^2}} y_i D_{A,co}^2 ,
\label{eq:si}
\end{equation}
where the integration is taken with respect to $n_{i}$, the number of galactic line emitters per cosmological comoving volume element. The factor $y_i$ is the derivative of the comoving radial distance with respect to the observed frequency, i.e. $y = d\chi/d\nu = \lambda_{i,rest} (1+z)^2/H(z)$, and $D_{A,co}$ is the comoving angular distance.

Finally, the shot noise component of the total line power spectrum---with the same units as the clustering term, namely, Jy$^2$ sr$^{-2}$ (Mpc h$^{-1})^{3}$---takes the form 
\begin{equation}
P_{i,i}^{shot}(z) = \int{\mathrm{d}n_{i} \left( \frac{L_{i}}{4\pi D_L^2} \right)^2 \left( y_i D_{A,co}^2 \right)^2}.
 \label{eq:pshot}
\end{equation}

\subsection{Calculating IR line volume emissivity}

The number density of line emitters and the line luminosity that appear in equations (2) and (3) can be derived by a variety of methods. In earlier papers on intensity mapping of molecular and fine-structure emission lines at high redshift (z $\gtrsim$ 6), one approach involved using the dark matter halo mass function in lieu of the line emitter density (and invoking a one-to-one correlation between halos and galaxies, which is reasonable at high redshifts). The line luminosity, in turn, could be scaled according to the star formation rate, which was related to halo mass via a proportionality constant comprised of factors describing the fraction of baryons available for star formation, as well as the dynamical timescale for star formation and a duty cycle for emission. While this approach is perhaps justified for the very early Universe (given the lack of information about the galaxy luminosity function at high redshift), the situation at later times is better understood; we make use of empirical constraints on the $z\sim1$ epoch from far-IR/submm number counts and observations of far-IR line emission in galaxies. 

\begin{figure}[t]
 \centering
 \includegraphics[width=0.45\textwidth]{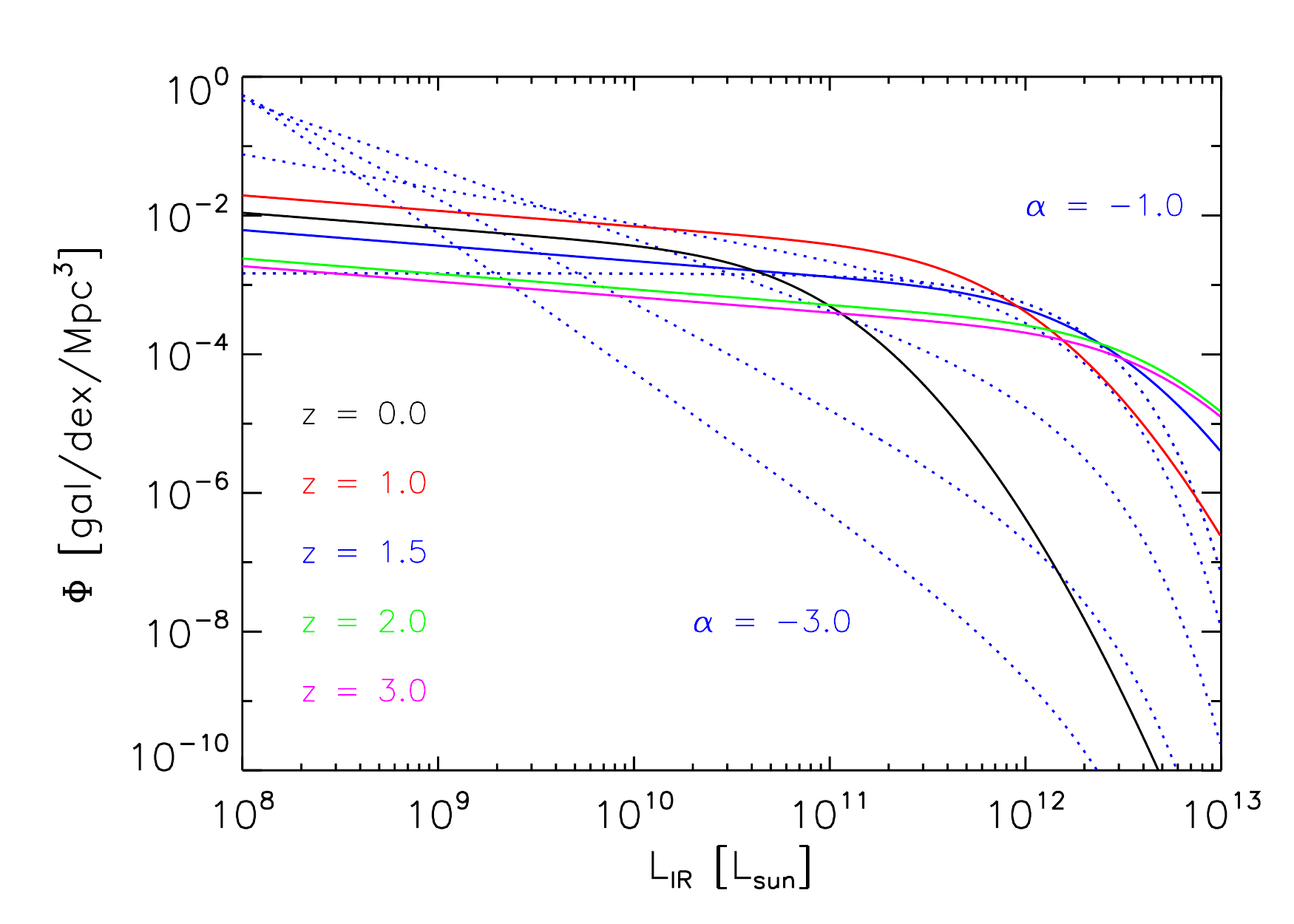}
\caption{B11 IR luminosity function computed at $z = 0.0, 1.0, 1.5, 2.0,$ and $3.0$ (solid black, red, blue, green, and magenta curves, respectively). Dotted blue curves represent Schechter-form luminosity functions---normalized such that the corresponding IR luminosity densities matches that of the B11 model---at $z = 1.5$ with faint-end slope (from top to bottom) $\alpha=-1.0, -1.5, -2.0, -2.5, -3.0$. Schechter functions with slopes steeper than $\alpha < -2.0$ are intended only for illustration.}
\label{fig:lum_funcs}
\end{figure}

We first employ the empirically-constrained, backwards-evolution model of the IR luminosity function $\Phi(L_{IR}, z)$ from \citet{bethermin11} (hereafter B11) to predict the number of galaxies with luminosity $L_{IR}$ at a given redshift in some comoving volume of the Universe per logarithmic luminosity interval, i.e., $\frac{\mathrm{d}N(L_{IR},z)}{\mathrm{d}V\mathrm{dlog_{10}}L_{IR}}$ or $\frac{\mathrm{d}n_{IR}}{\mathrm{dlog_{10}}L_{IR}}$:
\begin{equation}\begin{split}
\Phi(L_{IR}, z) &= \Phi_*(z) \left(\frac{L_{IR}}{L_*(z)}\right)^{1-\beta} \\ &\times\mathrm{exp}\left[-\frac{1}{2\xi^2}\mathrm{log}\left(1 + \frac{L_{IR}}{L_*(z)}\right)\right]
\label{eq:b11}
\end{split}
\end{equation}
In the above expression, $\beta$ and $\xi$ set the faint-end power law slope and the bright-end Gaussian width, respectively, of the luminosity function. When evaluating Equation~\ref{eq:b11} at different redshifts and luminosities, we use the best-fit parameters from B11 (cf. their Table 1), and so keep $\beta = 1.223$ and $\xi = 0.406$. The parameters $\Phi_*$ and $L_*$ follow a redshift evolution according to $\Phi_*(z) =  3.234\times{10}^{-3} \mathrm{gal/dex/Mpc^3}(1+z)^{r_{\Phi}}$ and $L_{*}(z) = 2.377\times10^{10} \mathrm{L}_{\odot} (1+z)^{r_L}$, where $r_{\Phi}$ and $r_{L}$ also have a redshift-dependence, given by
\begin{displaymath}
r_{\Phi} = \left\{
	\begin{array}{lr}
		0.774, &  z < z_{break,1} \\
		-6.246, &  z_{break,1} < z < z_{break,2} \\
		-0.919, &  z > z_{break,2}
	\end{array} 
	\right.
\end{displaymath}
\begin{displaymath}	
r_{L} = \left\{
	\begin{array}{lr}
		2.931, &  z < z_{break,1} \\
		4.737, &  z_{break,1} < z < z_{break,2} \\
		0.145, &  z > z_{break,2}
	\end{array} 
	\right.	
\end{displaymath}
The first break in redshift, $z_{break,1} = 0.879$, is a fitted parameter, whereas the second break is fixed at $z_{break,2} = 2.0$. In Figure~\ref{fig:lum_funcs}, we plot the B11 luminosity function at several different redshifts up to $z = 3$.

To convert the infrared luminosity to a line luminosity, we apply the relation for $L_{i}$ as a function of $L_{IR}$ provided by \citet{spinoglio12}. (Working directly from the IR luminosity function, we do not include the population of IR-dark or IR-faint sources that nevertheless may contribute bright emission in the far-IR fine-structure lines (cf. \citet{Riechers14}).) The fits in their paper were based on the diverse collection of ISO-LWS observations of local galaxies with luminosities between 10$^8$ and 10$^{13}$ L$_{\odot}$ from \citet{brauher08}. For example, we reproduce below the relation for [CII]:
\begin{equation}
\mathrm{log_{10}}L_{\mathrm{[CII]}} = (0.89 \pm 0.03) \mathrm{log_{10}}L_{\mathrm{IR}} - (2.44 \pm 0.07),
\end{equation}
indicating that [CII] is suppressed for higher luminosity systems. In general, the $L_i$-$L_{IR}$ relations can be written in the form
\begin{equation}
\mathrm{log_{10}}L_i = (A \pm \sigma_A) \mathrm{log_{10}}L_{\mathrm{IR}} - (B \pm \sigma_B),
\end{equation}
Slope, intercepts, and associated uncertainties described by the variables $A,B,\sigma_A,$ and $\sigma_B$ are summarized in Table~\ref{tab:spinoglio_relations} for a variety of IR lines.
\begin {table}[t]
\begin{center}
\caption {$L_{i}-L_{IR}$ relation variables from \citet{spinoglio12}}\label{tab:spinoglio_relations} 
\begin{tabular}{ l c c c c}
\hline \hline
Line $i$ & A & $\sigma_A$ & B & $\sigma_B$ \\
\hline
\textrm{[CII]158}$\mu$m & 0.89 & 0.03 & 2.44 & 0.07\\
\textrm{[NII]122}$\mu$m & 1.01 & 0.04 & 3.54 & 0.11 \\ 
\textrm{[OI]63}$\mu$m & 0.98 & 0.03 & 2.70 & 0.10 \\
\textrm{[OIII]88}$\mu$m & 0.98 & 0.10 & 2.86 & 0.30 \\
\textrm{[OIII]52}$\mu$m & 0.88 & 0.10 & 2.54 & 0.31 \\
\textrm{[SiII]35}$\mu$m & 1.04 & 0.05 & 3.15 & 0.16 \\
\textrm{[SIII]33}$\mu$m & 0.99 & 0.05 & 3.21 & 0.14 \\
\textrm{[SIII]19}$\mu$m & 0.97 & 0.06 & 3.47 & 0.20 \\
\textrm{[NeII]13}$\mu$m & 0.99 & 0.06 & 3.26 & 0.20 \\
\textrm{[NeIII]16}$\mu$m & 1.10 & 0.07 & 3.72 & 0.23 \\
\hline 
\end{tabular}
\end{center}
\end{table}

The choice of using local $L_i$-$L_{IR}$ relations for our study of $z\sim1$ emitters may be unrealistic due to findings that suggest the so-called ``deficit" in [CII] and other far-IR lines evolves with redshift such that the high-z counterparts to local systems do not exhibit suppressed far-IR line emission. The local IR relations then can be interpreted as underestimating emission of the fine-structure lines, since we likely overestimate the deficiency in the higher redshift, high luminosity systems of our model. While there are undeniably a number of uncertainties with the combined B\'{e}thermin-Spinoglio model, a simple extrapolation from the \citet{hb06} star formation history clearly brackets our predicted [CII] intensity at the relevant redshifts, and so we adopt it as our fiducial model throughout this paper. In Section 4, however, we explore variations in the shape of the IR luminosity function and consider an alternative line-to-IR luminosity ratio (depicted as the dotted curves in Figure~\ref{fig:lum_funcs}). 

Next, it becomes possible to write the cosmic mean intensity and shot noise of the line, in units of Jy sr$^{-1}$,  as a function of redshift based on the B11 luminosity function and \citet{spinoglio12} $L_{i}-L_{\mathrm{IR}}$ relation as
\begin{equation} \label{eq:intensity}
\bar{S}_{i}(z) = \int \mathrm{dlog}L_{IR}  \Phi(L_{IR}, z) \frac{f_{i} L_{IR} }{4 \pi D_{L}^2} y D_{A,co}^2
\end{equation}
\begin{equation} \label{eq:pshotlum}
P_{i,i}^{shot}(z) = \int \mathrm{dlog}L_{IR}  \Phi(L_{IR}, z) \left(\frac{f_{i} L_{IR}}{4 \pi D_{L}^2} y D_{A,co}^2\right)^2
\end{equation}
where the limits of integration are over the full range of expected IR luminosities, i.e. 10$^8$ to 10$^{13}$ L$_{\odot}$, and $f_{i}$, i.e. $\frac{L_{i}(L_{IR})}{L_{IR}}$, is the fraction of IR luminosity emitted in line $i$, as computed from equation (3). In other words, we have written $\bar{S}_{i}$ and $P_{i,i}^{shot}(z)$ as the first and the second moments of the luminosity function. 
\begin{figure}[t]
 \centering
 \includegraphics[width=0.45\textwidth]{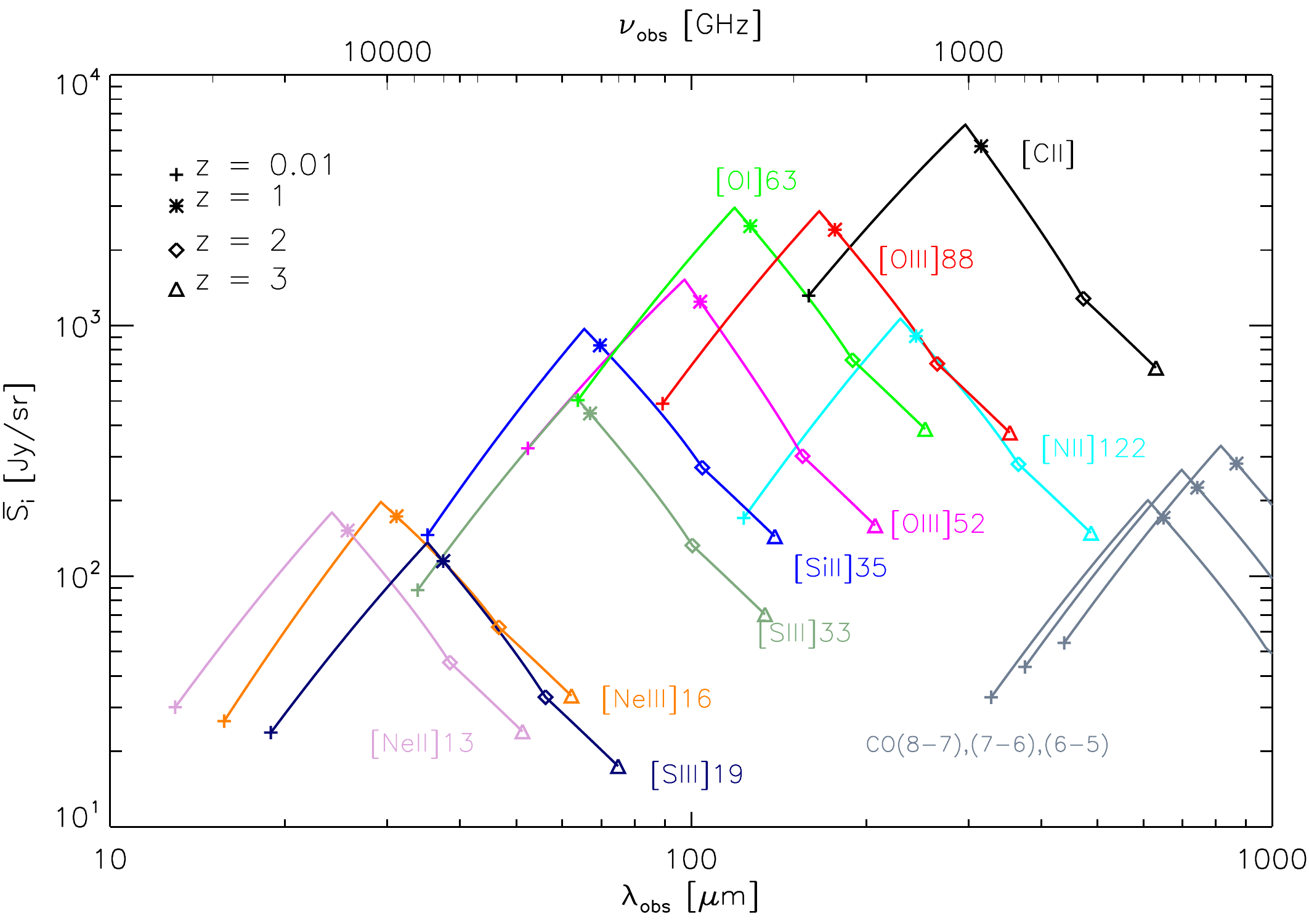}
\caption{Intensity of fine-structure line emission as a function of observed wavelength for the empirical model based on the B11 luminosity function. Intensities of CO lines, which are not included in the IR luminosity relations from \citet{spinoglio12}, have been estimated using a luminosity scaling provided by \citet{carilli11} for CO(1-0) and the relative intensities of the higher-J lines in \citet{bothwell13}.}
\label{fig:interlopers}
\end{figure}
The resulting mean intensities for a variety of far-IR lines are plotted in Figure~\ref{fig:interlopers} as a function of redshift and observed wavelength. $\bar{S}_{i}$ vs $\lambda_{obs}$ can be interpreted as identifying the dominant source of fluctuations, according to our model, of a given wavelength. As a specific example, if the target line of an observation is [OI]63$\mu$m at $z = 1$, it is necessary to distinguish between the target line and interlopers like [OIII]88$\mu$m from $z=0.4$ and [OIII]52$\mu$m from $z=1.4$, which contribute power at the observed wavelength. \citet{visbal10} showed how the cross spectra can be used to differentiate between a target line and a contaminating line (or ``bad line", in their words), since emitters at different redshifts will be spatially uncorrelated. For the observed wavelengths of [CII], however, it is apparent from Figure 2 that, with the exception of contributions from [OIII]88$\mu$m and CO(8-7) near [CII] at $z \sim 0.01$ and $z>2$, respectively, the [CII] line is relatively unaffected by interloper lines---a result of its luminosity and spectral isolation. It is for this practical reason, and for the astrophysical significance of [CII] mentioned in the Introduction, that we focus the remainder of this paper largely on [CII] emission.

\subsection{[CII] Luminosity Functions and Expected Power Spectra}

\begin{figure}
\centering
\includegraphics[width=0.45\textwidth]{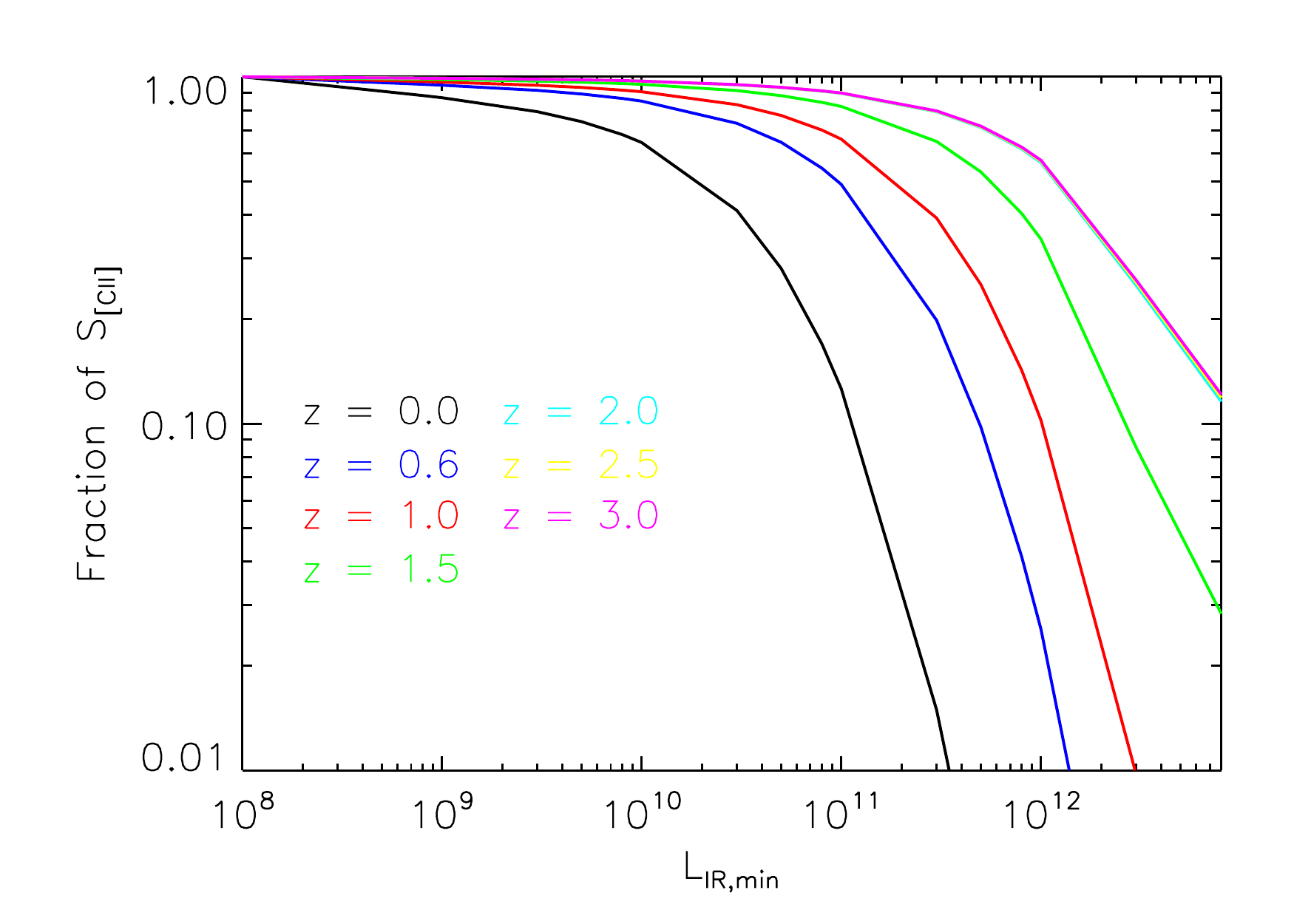}
\caption{The fraction of total [CII] mean intensity as a function of lower limit in the luminosity function. Different color curves represent different redshifts, as labeled on the plot.}
\label{fig:frac_cii_b11_lirmin}
\end{figure}
As laid out in Equations~\ref{eq:pclust} and ~\ref{eq:si}, $P_{\textrm{[CII],[CII]}}^{clust}$ is sensitive to intensity fluctuations from the full range of normal ($L_{IR} < 10^{11}$ L$_{\odot}$) to ULIRG-class ($L_{IR} > 10^{12}$ L$_{\odot}$) systems because its amplitude is proportional to the mean line intensity, squared. The information contained in a power spectrum of individually detected galaxies is, in contrast to the line intensity mapping approach, necessarily limited to galaxies which are above a certain detection threshold, or $L_{IR,min}$. Figure~\ref{fig:frac_cii_b11_lirmin} shows the integrated luminosity functions for [CII] in our model, which gives a sense of the depth that a galaxy survey must reach in order to completely probe the full integrated [CII] emission, i.e. all of $\bar{S}_i$. In this section, we examine the role of the various luminosity ranges on the amplitude of the observed [CII] power.  

\begin{figure}
\centering
\includegraphics[width=0.45\textwidth]{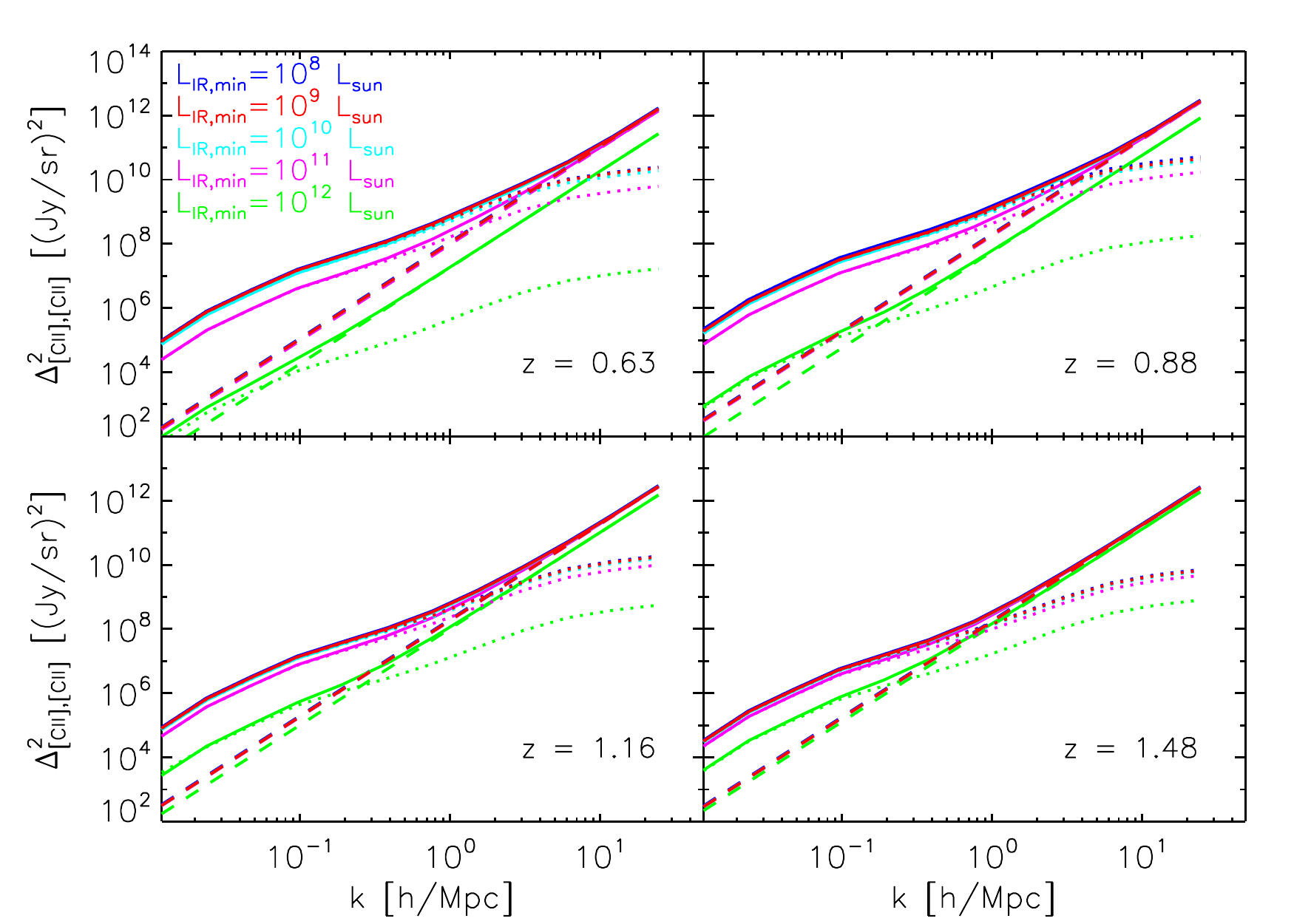}
\caption{Predicted [CII] autocorrelation power spectra from $z = 0.63$ to $z = 1.48$. Blue, red, cyan, magenta, and green curves represent the power spectrum computed with a lower limit in the luminosity function corresponding to $10^8$, $10^9$, $10^{10}$, $10^{11}$, and $10^{12}$ L$_{\odot}$, respectively. Dotted curves indicate power from clustering (including contributions from linear and nonlinear terms), and dashed curves indicate the contribution from shot noise power.}
\label{fig:pcii_lirmin}
\end{figure}

Power spectra at four representative redshifts ($z = 0.63, 0.88, 1.16$, and $1.48$) comprised of the sources above a few different survey depths, or  $L_{IR, min}$, are represented by Figure~\ref{fig:pcii_lirmin}. (Note that we use  $\Delta_{\textrm{[CII],[CII]}}^2 = k^3 P_{\textrm{[CII], [CII]}}(k)/(2\pi^2)$ when plotting the power spectrum. In this notation, the factor $k^3$ cancels out the volumetric units of $P_{\delta,\delta}(k,z)$ and the integral of $\Delta_{\textrm{[CII],[CII]}}^2$ over logarithmic k bins is equal to the variance in real space.) At these redshifts, the average linear bias has been assumed to take the observationally-motivated values of $\bar{b} = 2.0, 2.3, 2.6$, and $2.9$, though, in general, the bias will likely depend on the galaxy luminosity provided that luminosity is correlated with halo mass. In this Figure, we see the clustering amplitude decrease as the IR detection threshold is raised from 10$^{8}$ L$_{\odot}$ to 10$^{12}$ L$_{\odot}$.  (Note that the reduction in the clustering amplitude is precisely the square of the factor of reduction in $\bar{S}_{\textrm{[CII]}}$ plotted in Figure~\ref{fig:frac_cii_b11_lirmin}.) The level of decrease in clustering power as a result of raising $L_{IR,min}$ is most dramatic at the lower end of the redshift range of interest, when the luminosity function is represented mostly by normal galaxies and LIRGs. As ULIRGs rise to dominate the IR luminosity function at $z\sim1.5$, the amplitude of the clustering component of $P_{\textrm{[CII],[CII]}}(k,z)$ becomes relatively robust up to $L_{IR,min} \sim 10^{11}$ L$_{\odot}$, implying that a large fraction of the fluctuations are captured at this depth; we infer from Figure~\ref{fig:frac_cii_b11_lirmin} that, at $z = 1.48$, individually resolving galaxies at a depth of $6\times10^{11}$ will recover half of the [CII] light, at which point the remaining power of unresolved fluctuations is 25\% according to our model. For redshifts $z=0.63, 1.16$ and $3.0$, the corresponding depths to observe half-light are $\sim 10^{11}$, $2\times10^{11}$, and $10^{12}$ L$_{\odot}$, respectively. 

\section{The [CII] Power Spectrum}

\subsection{Observational Sensitivity to the Power Spectrum}

\begin{figure*}[t]
\centering
\begin{tabular}{cc}
\includegraphics[width = 0.49\textwidth]{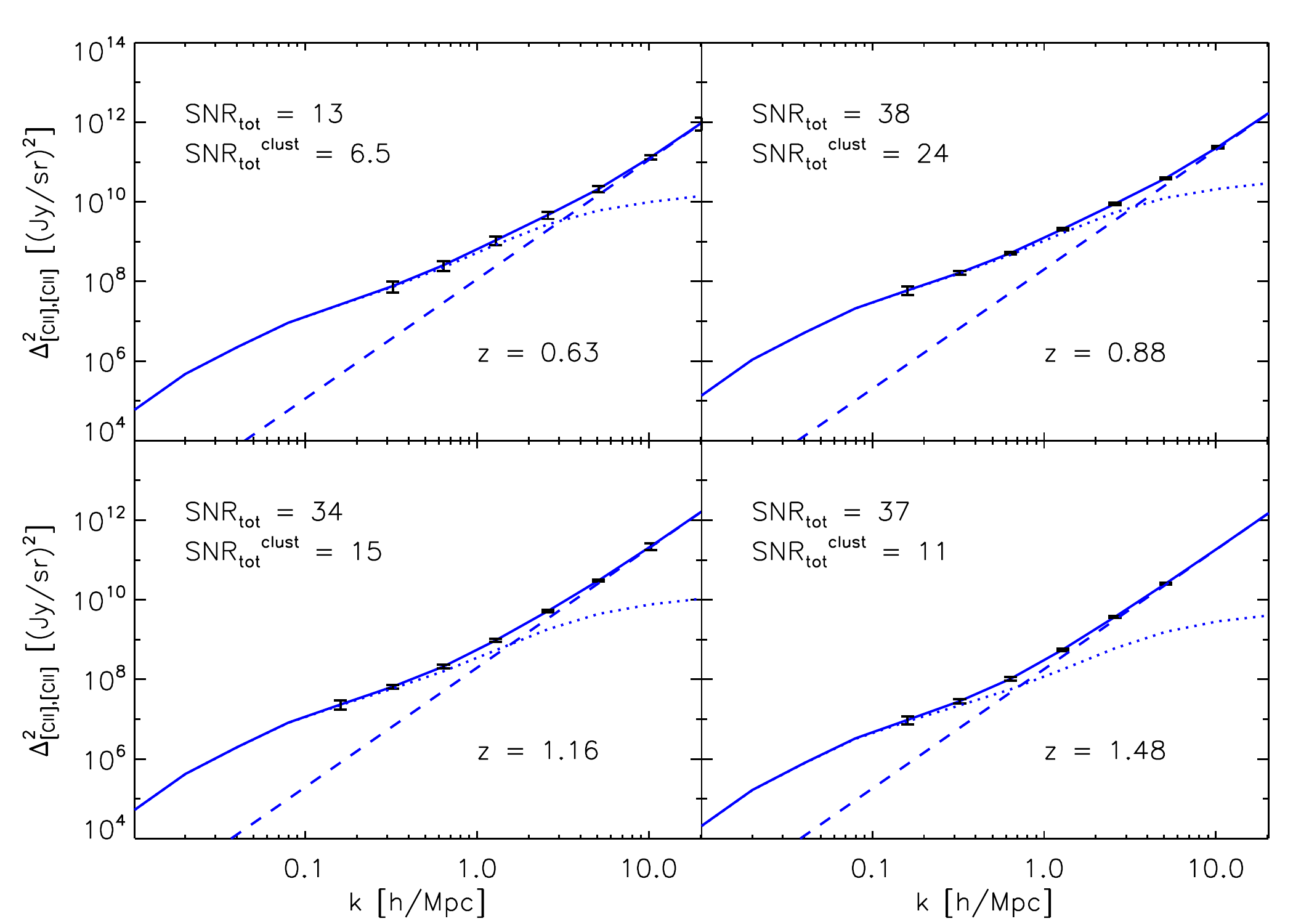} &
\includegraphics[width = 0.49\textwidth]{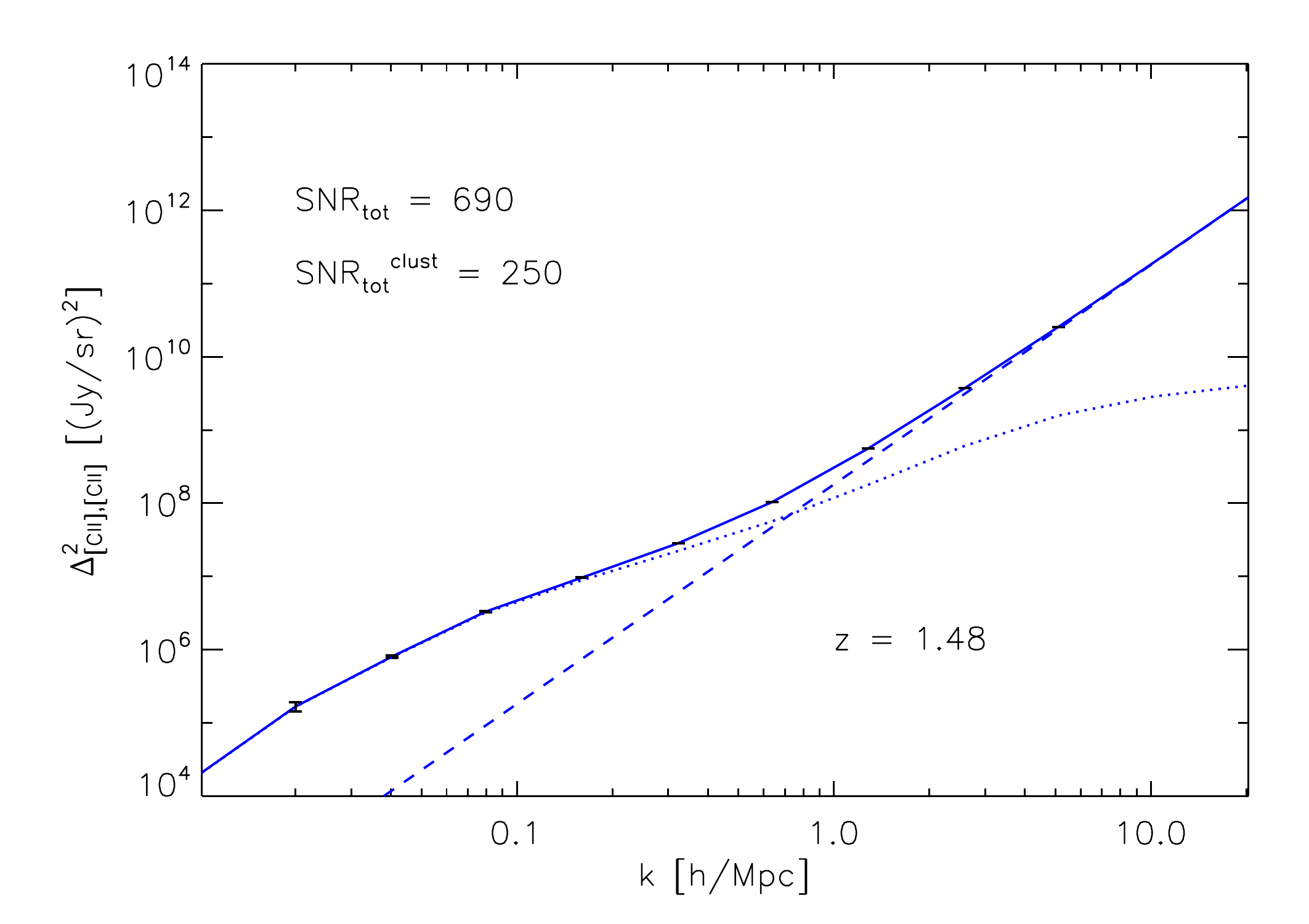} \\
\end{tabular}
\caption{Left panel: Predicted [CII] power spectra with error bar estimates from $z$ = 0.63 to $z$ = 1.48 for the fiducial balloon experiment, and with a total observing time of 450 hours. Dotted curves indicate power from clustering (including contributions from linear and nonlinear terms), and dashed curves indicate the contribution from shot noise power. Right panel: [CII] power spectrum expected at $z=1.48$ with error bar estimates for the fiducial cryogenic satellite experiment.}
\label{fig:pcii_zall}
\end{figure*}
We present in this section an assessment of detectability of the [CII] power spectrum. In order to quantify the observational sensitivity, we consider realistic experimental platforms with uninterrupted wavelength coverage in the redshift range of interest, namely, from 240 to 420 $\mu$m. This range is further divided into four bands to enable measuring redshift evolution in the signal. The width of each band has been set to span a redshift range of $\frac{\Delta{z}}{z_{center}} = 0.25$ to ensure there is no significant cosmological evolution within the band. Fiducial experimental parameters are summarized in Table ~\ref{tab:ExpParams}, though we explore the effect of varying $D_{ap}$ and $A_{survey}$ on the signal-to-noise Ratio (SNR).

To define the survey depth, we adopt the quantity 
\begin{equation}
f_{err} \equiv \frac{\sigma_N}{\sqrt{t_{obs}^{vox}} \bar{S}_i}
\label{eq:ferr}
\end{equation}
which we call the fractional error. It is simply the inverse of the SNR on the mean intensity in a single voxel. Here $\sigma_N$ is the instrument sensitivity (noise equivalent intensity, or NEI, in units of Jy sr$^{-1}$ s$^{1/2}$, $\bar{S}_i$ is the mean intensity and $t_{obs}^{vox}$ is the observing time per voxel. (We take $i=\mathrm{[CII]}$ while the equations remain generally applicable to any line.) Error bar estimates and the total SNR for the power spectrum are calculated by assuming a spectrally flat noise power spectrum, so that the noise power in each voxel, $P_{N}$, is written as
\begin{align}
P_N & = \sigma_N^2 \frac{V_{vox}}{t_{obs}^{vox}} \\
& = \left(f_{err} \bar{S}_i \right)^2 V_{vox} \nonumber
\end{align}
where $V_{vox}$ is the volume of a voxel. The voxel volume is the product of pixel area, $A_{pix}$ (in units of comoving Mpc$^2$ h$^{-2}$), and the line of sight distance along a spectral channel, $\Delta$r$_{los}^{vox}$ (Mpc h$^{-1}$). $A_{pix}$ depends on the telescope aperture and observed wavelength according to $A_{pix} = (\lambda_{i,obs}/D_{ap}\times D_{A})^2$.

The variance of a measured $k$, $\sigma^2(k)$, is then written as
\begin{equation}
\sigma^2(k) = \frac{\left({P_{i,i}(k) + P_N(k)}\right)^{2}}{N_{modes}(k)},
\label{eq:variance}
\end{equation}
where $N_{modes}$ is the number of wavemodes that are sampled for a given $k$ bin of some finite width $\Delta$log(k). (We have chosen $\Delta$log(k) = 0.3 for this analysis.) We restrict the mode counting to the upper-half plane in $k$-space, so as not to overestimate the number of independent modes sampled. 

The total SNR, in turn, is calculated from the expression 
\begin{equation}
SNR_{tot} = \sqrt{\sum_{bins} \left(\frac{P_{i,i}(k)}{\sigma(k)}\right)^2}
\label{eq:snrtot}
\end{equation}

The expected [CII] power spectrum, with corresponding predictions for SNR, at the same redshifts from Figure~\ref{fig:pcii_lirmin} are shown in Figure~\ref{fig:pcii_zall}. In calculating the power spectrum sensitivity for these power spectra, the two lowest line-of-sight modes and the lowest transverse mode are not included, since these modes will likely be compromised by the necessity of continuum foreground subtraction and beam-differencing in the fluctuation analysis. (The exact effect of continuum subtraction will need to be modeled via simulation.)
\begin {table*}[h]
\begin{center}
\caption {Parameters for Envisioned Experimental Platforms} \label{tab:ExpParams} 
\begin{tabular}{ l c c c c}
\hline \hline
$D_{ap}$ (m) &\multicolumn{4}{c}{2.5} \\
$R = \lambda_{obs}/\Delta\lambda$ & \multicolumn{4}{c}{450} \\
Number of Spectral Channels & \multicolumn{4}{c}{64} \\
$N^{spatial}_{instr}$ (instantaneous spatial pixels) & \multicolumn{4}{c}{25} \\
$t_{obs}^{survey}$ (hr)&\multicolumn{4}{c}{450}\\
\hline
$z_{cen}$ for [CII] & 0.63 & 0.88 & 1.16 & 1.48 \\
Wavelength Range ($\mu$m) & 240-276 & 276-317 & 317-365 & 365-420 \\
$V_{voxel}$ (Mpc$^3$ h$^{-3}$) & 0.36 & 0.81 & 1.59 & 2.87 \\
$A_{pix}$ (Mpc$^2$ h$^{-2}$) & 0.044 & 0.096 & 0.19 & 0.35 \\
$\Delta$r$_{los}^{vox}$ (Mpc h$^{-1}$) & 7.8 & 7.8 & 7.7 & 7.5 \\
$\bar{S}_{\mathrm{[CII]}}$ (Jy sr$^{-1}$) & 4.56 $\times 10^3$ & 6.33 $\times$ 10$^3$ & 4.05 $\times 10^3$ & 2.55 $\times 10^3$ \\
\hline
\hline
 & \multicolumn{4}{c}{\emph{Atmospheric Balloon}}\\
$A_{survey}$ (deg$^2$) & 1 & 1 & 1 & 1 \\ 
$\sigma_N$ (10$^7$ Jy sr$^{-1}$ sec$^{1/2}$) & 3.4 & 2.1 & 1.5 & 1.0  \\
Line Sensitivity, $S_\gamma$ (10$^{-18}$ W m$^{-2}$ sec$^{1/2}$) & 15.8 & 11.3 & 9.20 & 7.10 \\
$f_{err}$ & 160 & 63 & 61 & 56 \\
\hline
 & \multicolumn{4}{c}{\emph{Cryogenic Satellite}} \\
$A_{survey}$ (deg$^2$) & 1,000 & 1,000 & 1,000 & 1,000 \\ 
$\sigma_N$ (10$^7$ Jy sr$^{-1}$ sec$^{1/2}$) & 0.030 & 0.034 & 0.039 & 0.043 \\
Line Sensitivity, $S_\gamma$ (10$^{-18}$ W m$^{-2}$ sec$^{1/2}$) & 0.139 & 0.185 & 0.240 & 0.306 \\
$f_{err}$ & 45 & 32 & 50 & 77 \\
\hline
\end{tabular}
\end{center}
\end{table*}
Table~\ref{tab:ExpParams} shows our instrument concepts.  We specify a 25-beam grating spectrometer covering the 240-420 $\mu$m band, each with 64 $R=450$ spectral channels operating near the photon background limit, illuminated with a 2.5-meter telescope. We consider a balloon experiment for which the photon background is due to 1\% emissivity in the atmosphere (a conservative average value) and 4\% in the telescope. A 450 hour integration (as might be obtained in a long duration balloon flight) over the 1 square degree with this system results in the $\sigma_N$, $f_{err}$, and line sensitivity values tabulated. We also consider a similar instrument on a cryogenic space-borne platform. The sensitivity in this case is obtained by specifying a detector sensitivity which is equal to the photon background noise, so that the quadrature sum is $\sqrt{2}$ times the photon noise. The photon background is taken to be due to the combination of zodiacal light, galactic dust, and a 6-K telescope telescope with 4\% emissivity. This is an optimized instrument with advanced detectors -- it is similar to the best case of the proposed BLISS instrument for SPICA (e.g., see \citet{bradford12}). As the tabulated depths indicate, the space-borne system is much more sensitive. Nevertheless, the balloon-borne experiment is capable of measuring the power spectrum with good sensitivity, and all error bars in this paper are based on the 450-hour balloon experiment, unless otherwise noted. 

We find that the total power spectrum, including power from both shot noise and clustering, is observable using the balloon platform with $\textrm{SNR}>10$ at all examined redshifts; the clustering power, in turn, can be detected with $\textrm{SNR}>10$ in the redshift range from $z = 0.88-1.48$. From space, it becomes feasible to survey larger areas ($\sim$1,000 deg$^2$) and maintain high SNR on the order of 100. (See Figure~\ref{fig:pcii_zall} for calculated SNRs.) 
\begin{figure}[h]
\centering
\begin{tabular}{c}
\includegraphics[width=0.40\textwidth]{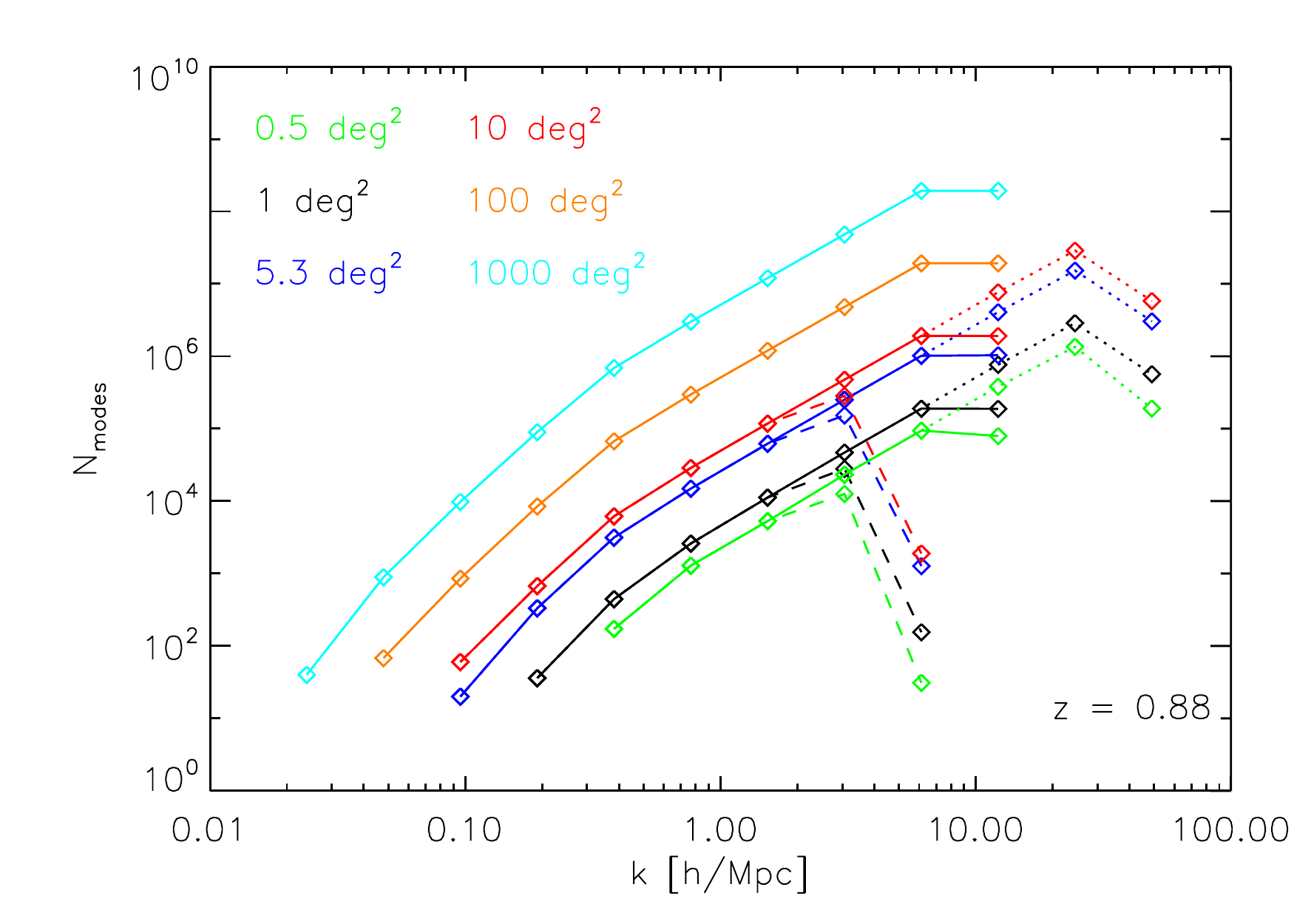} \\
\includegraphics[width=0.40\textwidth]{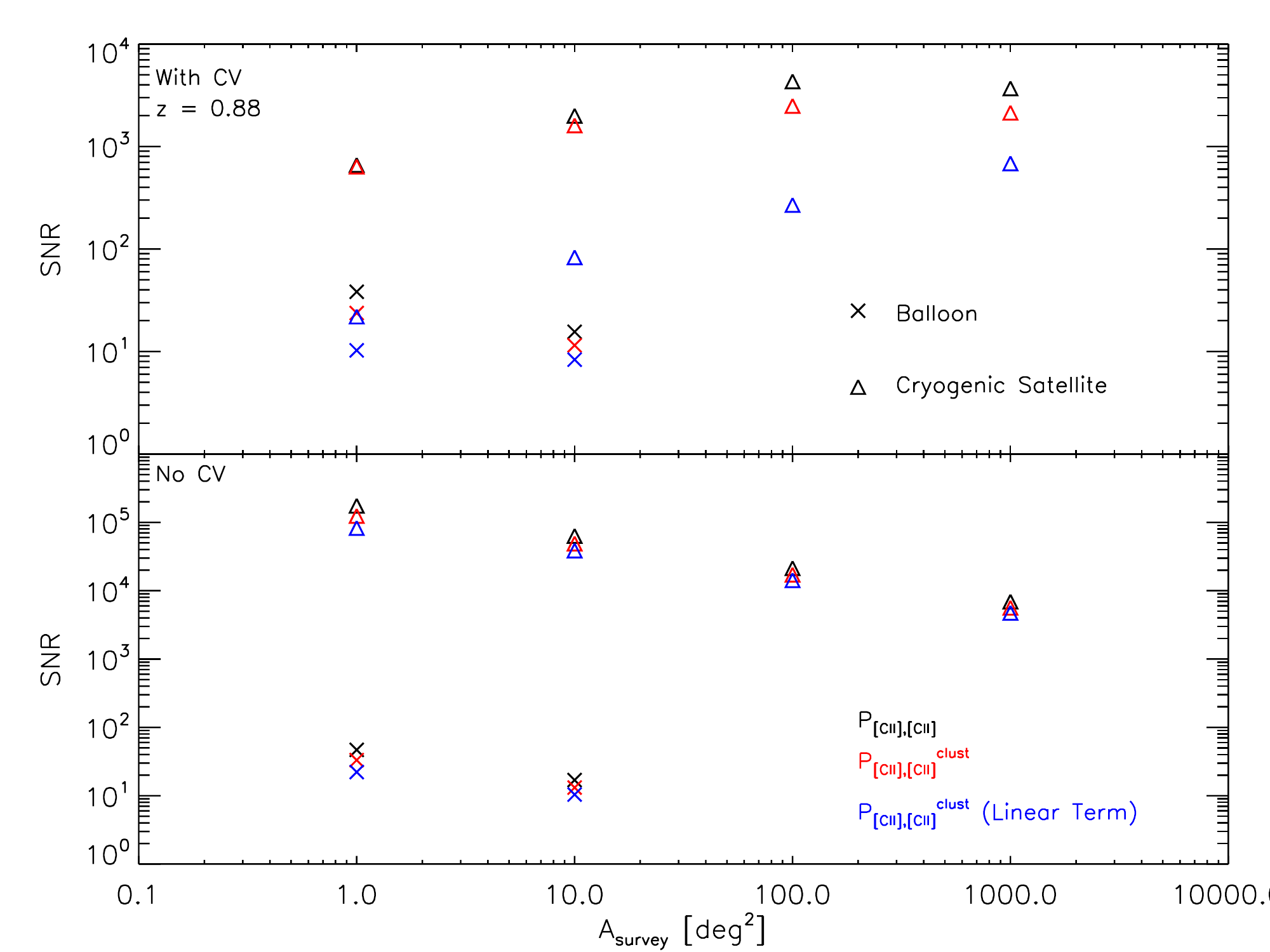}\\
\end{tabular}
\caption{Top panel: Number of modes as a function of $k$ at $z=0.88$ for different survey areas. Telescopes with apertures yielding 0.1, 1, and 10 times the fiducial $V_{vox}$ are shown as the dotted, solid, and dashed lines, respectively. Note that the decline in $N_{mode}$ at high-$k$ is an artifact of our method of counting modes on a pixellated, finite volume, where voxels are rectangular in shape, with transverse dimension matching the instrumental beam size and line-of-sight dimension matching the spectral resolution. Bottom panel: SNR on the total power spectrum (black), clustering power spectrum (red), and the linear portion ($k\lesssim0.1$ h/Mpc) of the clustering power spectrum (blue) with and without error from cosmic variance. Values for the balloon and cryogenic satellite experiments described in the text are designated with crosses and triangles, respectively.}
\label{fig:nmode_snr_asurvey}
\end{figure}
In the top and bottom panels of Figure~\ref{fig:nmode_snr_asurvey} we examine the effect of changing the survey area and telescope aperture on accessible wavemodes and SNR, where the number of modes has been plotted as a function of $k$, and SNR has been plotted as as a function of survey area. Our fiducial survey area of $A_{survey} = 1.0$ deg$^2$ for the balloon experiment is optimal for measuring as many large scale ($k\lesssim0.1$ h/Mpc) modes as possible with highest SNR in each $k$-bin, as illustrated in the lower panels of Figure~\ref{fig:nmode_snr_asurvey}. In this figure observing time is fixed, so the total SNR increases with survey area when modes are cosmic variance dominated---as in the case for the cryogenic satellite experiment---and decreases when modes are noise-dominated---as in the balloon experiment. When cosmic variance is not included, larger areas merely translate to lower integration time (i.e., greater noise) per voxel, and SNR decreases. The lack of significant change in SNR when including or excluding cosmic variance in the error budget for the balloon experiment indicates that the survey is not cosmic-variance limited. We do not consider surveys with areas less than a square degree because this prohibits measurement of power on large physical scales (cf. top panel of Figure~\ref{fig:nmode_snr_asurvey}).

To better demonstrate how the observational parameters drive the behavior of SNR, we rewrite $P_N$ in terms of the parameters from Table 1 (where the units of $A_{survey}$ have been converted to physical area in units of Mpc$^{2}$ h$^{-2}$) giving 
\begin{equation}
\begin{split}
P_N& = \left(\sigma_N^2 A_{pix} \Delta r_{los}^{vox}\right) / \left({\frac{t_{obs}^{survey}}{n_{beams}/N_{instr}^{spatial}}}\right) \\
& = \left(\sigma_N^2 A_{pix}\Delta r_{los}^{vox}\right) /  \left(\frac{t_{obs}^{survey} N_{instr}^{spatial}}{A_{survey}/A_{pix}}\right)\\
& = \sigma_N^2 \frac{\Delta r_{los}^{vox} A_{survey}}{t_{obs}^{survey} N_{instr}^{spatial}}
\end{split}
\label{eq:pnoise}
\end{equation}
In this form, it becomes apparent that---with fixed number of spatial pixels, spectral resolution, and total observing time---the only factor driving up the amplitude of noise power is the survey area; the effect of increasing aperture only allows access to higher wavenumbers, which is important for subtracting the shot noise from the total power to reveal the clustering.

\subsection{Measuring Line Luminosity Density Over Cosmic Time}

As noted above, intensity mapping is naturally sensitive to the full range of galaxy luminosities through the mean intensity, which is imprinted in the linear (2-halo) clustering term. Shot noise must be accurately subtracted, and this should be straightforward given the high SNR in the shot-noise dominated k bins (Figure~\ref{fig:pcii_zall}).  Next, per Equation~\ref{eq:pclust}, a measurement of the clustering power in the line emission directly constrains the product $\bar{S}_{i}^2 \bar{b}_i^2$. To extract $\bar{S}_i$, it is necessary to divide out $P_{\delta,\delta}(k,z)$ and  $\bar{b}_{\textrm{[CII]}}^2(z)$. The confidence with which these are \emph{a priori} known quantities becomes lower as $k$ increases. For example, the 1-halo power spectrum for DSFGs appears to be dependent on the IR luminosity of the contributing sources \citep{viero13}, indicating the need to map sufficiently wide areas that access $k$ modes where the power is largely independent of the level of 1-halo power. In the case of the galaxy bias, measurements of the angular dependence of the clustering can, in principle, be used to independently solve for $\bar{b}_i$ via the anisotropy in the angular power spectrum induced by redshift space distortions, as suggested in \citet{lidz11}. 

\begin{figure}[h]
\centering
\includegraphics[width=0.45\textwidth]{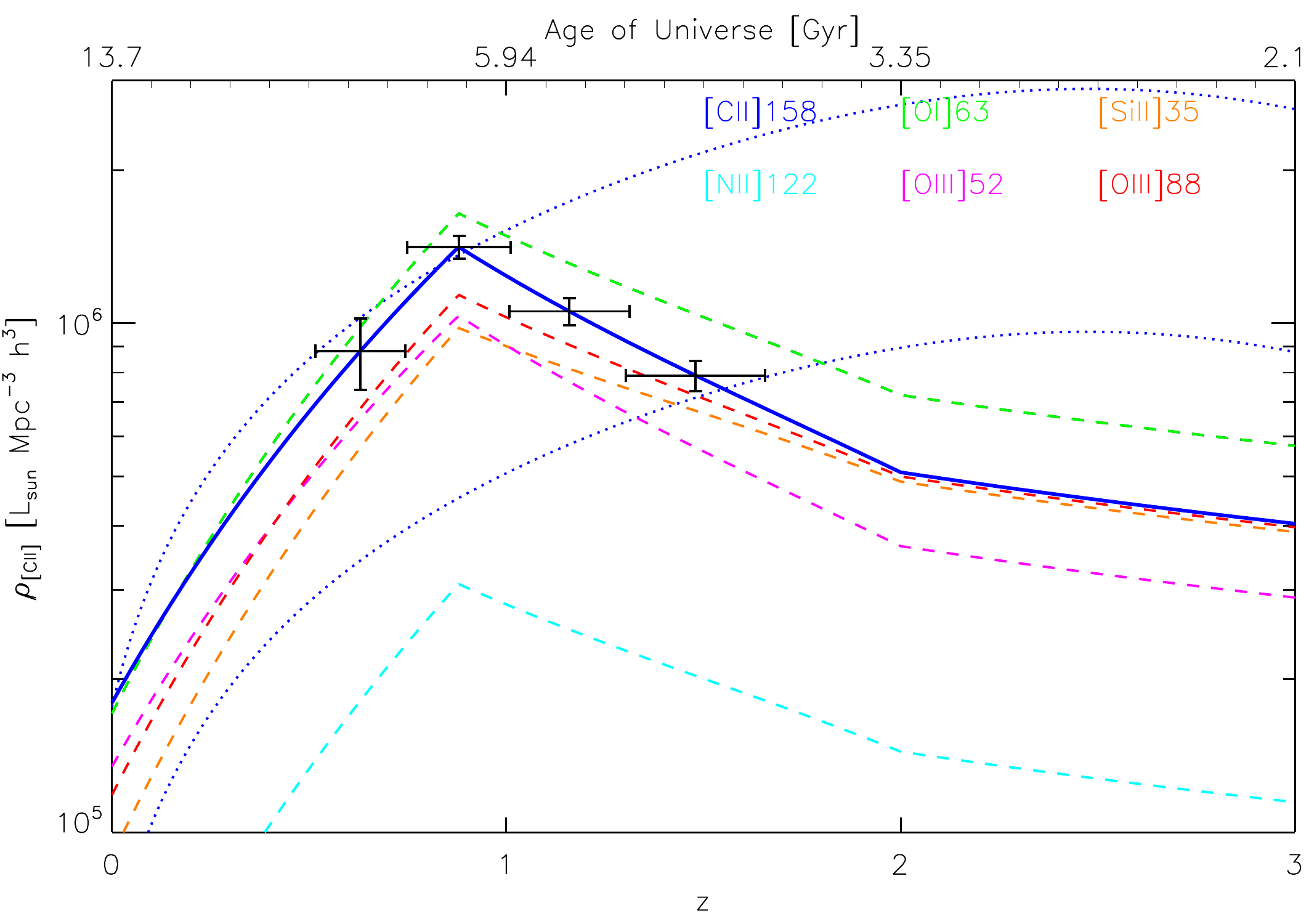}
\caption{Error bar estimates on $\rho_{\textrm{[CII]}}$, as measured by the fiducial balloon experiment, at redshifts $z$ = 0.63, 0.88, 1.16, and 1.48. Errors in $z$ correspond to the redshift space spanned by the spectrometer bandwidth. The solid blue curve is the underlying, fiducial B\'{e}thermin-Spinoglio model for [CII] luminosity density. The luminosity density of other bright IR lines, also from the fiducial model, are shown as the dashed colored curves, and the dotted curve is an estimate for $\rho_{\textrm{[CII]}}$ based on the fit to SFRD($z$) provided by \citet{hb06}, where we have used constant ratios of $L_{\textrm{[CII]}}$ to $L_{IR}$ equal to 0.001 (bottom curve) and 0.003 (top curve) to convert from IR luminosity density to [CII] luminosity density. Note that one can distinguish the different cosmic [CII] emission histories with the fiducial balloon experiment.}
\label{fig:scii_z}
\end{figure}

Returning to Figure~\ref{fig:nmode_snr_asurvey} (top panel), we see that, for the purpose of measuring $\bar{S}_{\textrm{[CII]}}$ with the fiducial survey of 1 deg$^2$ with the balloon experiment, there are two $k$ bins ($k = 0.16$ and 0.32 h/Mpc) in which the 2-halo clustering accounts for at least 80\% of the total power. (A survey with 10 deg$^2$, also shown in Figure~\ref{fig:nmode_snr_asurvey}, is wide enough to have three $k$ bins available in the linear regime, but the sensitivity on the additional mode with $t_{obs}^{survey}$= 450 hours is marginal.) Thus, in considering the case of $A_{survey} = 1.0$ deg$^2$, we find that it is possible to measure the co-moving [CII] luminosity density, $\rho_{\textrm{[CII]}}(z)$, in physical units of L$_{\odot}$ (Mpc/h)$^{-3}$,
\begin{align}
\rho_{\mathrm{[CII]}}(z) &= \int \mathrm{dlog}L_{IR}  \Phi(L_{IR}, z) f_{\mathrm{[CII]}}L_{IR} \\ &= \bar{S}_{\mathrm{[CII]}} 4\pi\lambda_{\mathrm{[CII]},rest}H(z),  \label{eq:rho_si_map}
\end{align}
within $\sim10\%$ accuracy from $z = 0.63$ to $z=1.48$, as depicted in Figure~\ref{fig:scii_z}. Here, the fractional uncertainty on $\rho_{\mathrm{[CII]}}(z)$ (or, equivalently, on $\bar{S}_{\textrm{[CII]}}(z)$ via the mapping described in Equation~\ref{eq:rho_si_map}) has been calculated according to standard error propagation as half the fractional uncertainty on $P_{\textrm{[CII],[CII]}}(k,z)$, so that the SNR on  $\bar{S}_{\mathrm{[CII]}}(z)$ is twice the SNR on the clustering power spectrum, $P_{i,i}^{clust}(k,z)$:
\begin{equation}
\mathrm{SNR \ on \ }\bar{S}_{\mathrm{[CII]}} = 2\times\sqrt{\sum_{\mathrm{linear\ k-bins\ only}} \left(\frac{P_{i,i}^{clust}(k)}{\sigma_{clust}(k)}\right)^2},
\label{eq:snrtot}
\end{equation}
where $\sigma_{clust}$ is merely the shot-noise subtracted version of Equation~\ref{eq:variance}, or, explicitly,
\begin{equation}
\sigma_{clust}(k) = \sqrt{ \frac{\left({P_{i,i}^{clust}(k) + P_N(k)}\right)^{2}}{N_{modes}(k)}}
\end{equation}
In Figure~\ref{fig:scii_z}, we also include, for comparison, an estimate for $\rho_{\textrm{[CII]}}(z)$ based on the analytic fit to SFRD($z$) provided by \citet{hb06} and flat ratios of $L_{\textrm{[CII]}}/L_{IR} = 0.001$ and 0.003. (For this purpose, we use the standard relation between SFRD and infrared luminosity described in \citet{kennicutt98}.)

The cryogenic satellite offers an unprecedented platform for quantifying the evolution of far-IR line emission in cosmological volumes over time, with fractional uncertainties on the order of a tenth of a percent at each redshift for the 1,000 deg$^2$ survey ($t_{obs}^{survey} = 450$ hr).

\section{Observational Strategy: Comparing Intensity Mapping with Traditional Galaxy Surveys}

\subsection{Probes of the mean line intensity}

Now let us turn to a question regarding the motivation for intensity mapping in general, as well as in the specific case of [CII] at the redshifts relevant to this study. Having identified the galaxy redshift surveys as an alternative method to measure the mean intensity of the line-emitting galaxy population and to measure the 3-D clustering power spectrum, it is natural to draw a comparison of the two approaches. 

The principal advantage of intensity mapping is that it naturally measures the aggregate emission per Equation~\ref{eq:intensity}, since the power spectrum depends on the integral of the [CII] luminosity function. Galaxy surveys always miss some of the light in the faintest galaxies, and this completeness problem is illustrated in Figure~\ref{fig:ngal_frac}. To make concrete comparisons in what follows we employ toy models for the infrared luminosity function (dotted curves in Figure~\ref{fig:lum_funcs}) written in the Schechter formalism---parametrized by the usual $\alpha$, $L_*$, and $\phi_*$---and normalize the total IR luminosity density according to B11 (cf. Appendix for details). We stress that these Schechter models are not intended to represent a real interpretation of the distribution of galaxies, but are merely helpful for illustrating the effect of the LF \emph{shape} on the relative usefulness of intensity mapping and traditional galaxy surveys. In converting the IR LF to a line luminosity function, we use, in addition to the \citet{spinoglio12} relation for $L_{\textrm{[CII}}/L_{IR}$, a conservative and flat line-to-IR luminosity ratio of $10^{-3}$, relegating the luminosity-dependence of this ratio (and any redshift evolution) as a second order effect. 

The line sensitivity, $S_{\gamma}$ (units of W m$^{-2}$ s$^{1/2}$), is the figure of merit for detecting an unresolved line in a point source, and we define individual detections at the $5\sigma$ level as having a flux above the instrumental noise in a voxel, i.e., above $5 \times \frac{S_{\gamma}}{\sqrt{t_{obs}^{vox}}}$. (In addition to instrumental noise, both Poisson fluctuations in the abundance of faint sources as well as the clustering of these sources may impact the ability to detect galaxies in the survey. However, we have explicitly checked that this ``confusion noise" is subdominant compared to instrumental noise for surveys considered in this work and do not consider this further here.) A convenient expression, which explicitly ties the minimum detectable line luminosity to a set of theoretical and experimental parameters, for the detection threshold can be written as
\begin{equation}
L_{i, min} = 5 \times f_{err} \rho_{i} V_{vox}, 
\label{eq:lirmin}
\end{equation}
Here, $f_{err}$ is the fractional error (Eq.~\ref{eq:ferr}) and $\rho_{i}$ is the comoving luminosity density of line $i$ at some $z$, or $L_*\phi_* \Gamma(2+\alpha, L/L_*)$ in the Schechter notation, so that equality holds between Equation~\ref{eq:lirmin} and the more conventional expression for the 5$\sigma$ detection threshold:
\begin{equation}
\frac{L_{i,min}}{4\pi D_L^2} \Leftrightarrow 5 \times \frac{S_{\gamma}}{\sqrt{t_{obs}^{vox}}}
\end{equation}
\begin{figure}[b]
\centering
\includegraphics[width=0.5\textwidth]{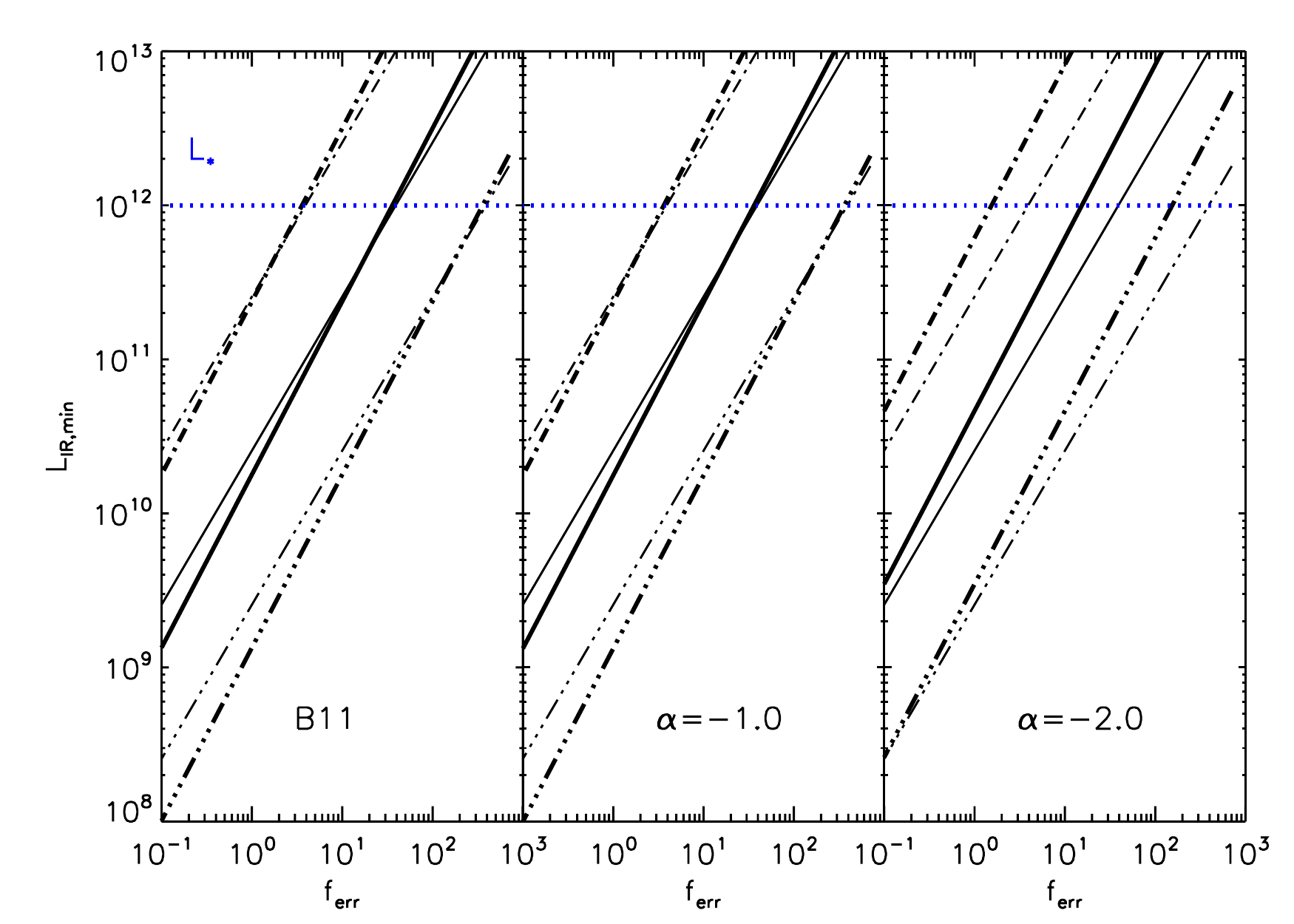}
\caption{IR depth as a function the fractional error. $L_{IR,min}$ refers to the minimum source luminosity that may be directly in the survey at 5$\sigma$ confidence. Results are plotted for the B11 (leftmost panel) model as well as the toy Schechter functions (remaining panels). Solid curves correspond to the fiducial aperture, $D_{ap}$ = 2.5 m. Dashed curves correspond to apertures scaled by a factor $\sqrt{\epsilon}$, where $\epsilon=10$ (triple-dot-dashed) and $\epsilon=0.1$ (dot-dashed). Thick curves correspond to our fiducial model for [CII] line intensity, based on Spinoglio fits, whereas thin curves denote the use of a constant ratio of $\frac{L_{\textrm{[CII]}}}{L_{IR}} = 10^{-3}$.}
\label{fig:lirmin_ferr}
\end{figure}

The survey depths $L_{IR,min}$ as a function of $f_{err}$, $V_{vox}$, and $\alpha$ are plotted in Figure~\ref{fig:lirmin_ferr}. Note that we are investigating the effect of changing telescope aperture, which only changes the transverse dimensions of $V_{vox}$.

Since the intensity mapping technique contains information in the power spectrum from sources below a given $S_{\gamma}$, we expect that regimes in which the majority of galaxies are too faint to be resolved are better-suited for intensity mapping observations than observations via the traditional galaxy survey. Inspection of Equation~\ref{eq:lirmin} yields that this scenario occurs for large voxels (or large beam sizes), large fractional errors, or steep luminosity functions where the bulk of the galaxy number density is comprised of galaxies with sub-$L_*$ luminosities. These three limiting cases for the fiducial square degree survey at $z=1.48$ are illustrated in Figures~\ref{fig:ngal_frac} and \ref{fig:snr_vs_ferr} for the experimental goals of measuring mean intensity and the clustering power spectrum, respectively.

As an example of the problem posed by steep luminosity functions for galaxy surveys aiming to measure the mean intensity, we refer to the top panel of Figure~\ref{fig:ngal_frac}. Here, we find that for LFs with $\alpha$ of -1.5 (not shown) or -2.0, the galaxy surveys detect only 30\% and $<1$\% of the total [CII] light in integrating to an $f_{err}$ of 10. Increasing the telescope aperture by a factor of $\sqrt{10}$ (shown as the triple-dot-dashed curves) boosts this fraction to 60\% in the case of $\alpha = -1.5$, but still recovers 10\% or less of the $\rho_{\mathrm{[CII]}}$ for $\alpha = -2.0$. 

\begin{figure}[h]
\centering
\includegraphics[width=0.5\textwidth]{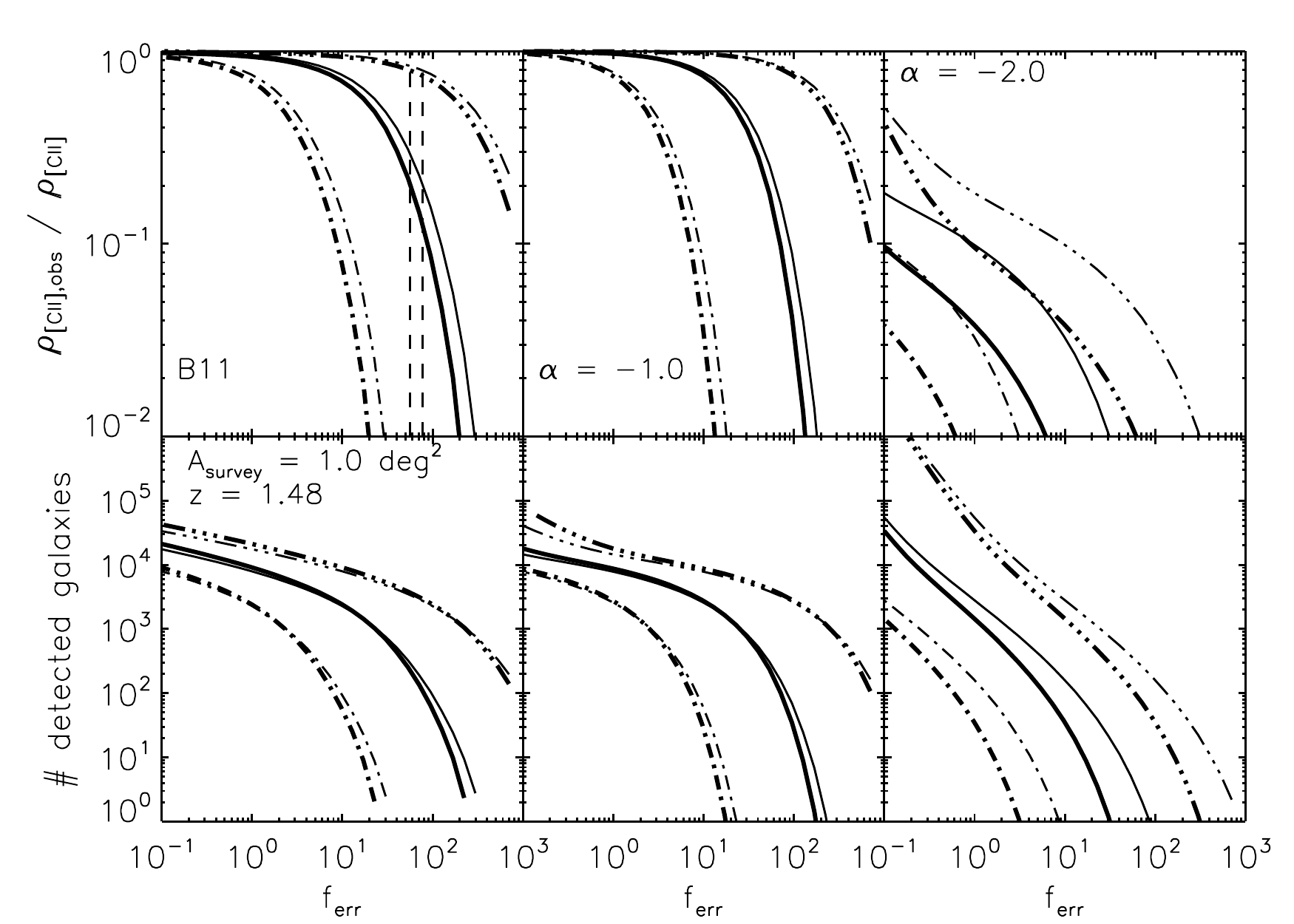}
\caption{Observed fraction of [CII] luminosity density as a function of survey time for the square degree field and the predicted number of [CII]-detected galaxies . Results are plotted for the B11 (leftmost panel) model as well as the toy Schechter functions (remaining panels). Solid curves correspond to the fiducial aperture, $D_{ap}$ = 2.5 m. Dashed curves correspond to apertures scaled by a factor $\sqrt{\epsilon}$, where $\epsilon=10$ (triple-dot-dashed) and $\epsilon=0.1$ (dot-dashed). Thick curves correspond to our fiducial model for [CII] line intensity, based on Spinoglio fits, whereas thin curves denote the use of a constant ratio of $\frac{L_{\textrm{[CII]}}}{L_{IR}} = 10^{-3}$. Reference values of $f_{err}$ for the fiducial balloon- and space-borne experiments are shown as dashed vertical lines.}
\label{fig:ngal_frac}
\end{figure}

The bottom row of Figure~\ref{fig:ngal_frac} breaks down the total emission in terms of the number of detectable galaxies. As is clear from comparison of panels in the top and bottom rows, a large sample of galaxies (of order 1,000 or greater) does not necessarily ensure an unbiased measure of the mean [CII] intensity. If, however, one extracts the aggregate, unresolved emission from [CII] via the intensity mapped power spectrum, one is essentially measuring $\frac{\rho_{\textrm{[CII]}, obs]}}{\rho_{\textrm{[CII]}}} = 1$ as soon as SNR on the linear clustering term of the power spectrum is sufficiently high, which was depicted in Figure~\ref{fig:scii_z}. 

Note that $f_{err} = 1$ allows the galaxy survey to reach a depth ($L_{IR,min} = 4\times10^{10}$ L$_{\odot}$ according to Figure~\ref{fig:lirmin_ferr}) corresponding to 90\% of the total [CII] light at $z=1.48$ for the B11 model, as shown in the top and leftmost panel of Figure~\ref{fig:ngal_frac}. A survey to this depth therefore might offer a means to extract the mean intensity by simply integrating the luminosity function. Such a low fractional error, however, requires either very low instrument noise or very long integration times---roughly $10^4$ hours for the fiducial balloon-borne instrument when observing a square degree field, for instance. (We refer the reader to Table~\ref{tab:ferrconvert} for the conversions between $f_{err}$ and integration time per voxel for the fiducial balloon experiment, as well as for the cryogenic satellite experiment.)

\subsubsection{Comparison with small-beam ground-based surveys}

Observations from the ground will, of course, lack redshift coverage as they are restricted to known atmospheric windows, yet we examine more closely the ability of ground-based facilities---current and planned---to constrain the mean [CII] intensity with individual detections.

For observations with an ALMA pencil beam survey at $z = 1.2$ (860 GHz, or roughly the central frequency of Band 10), the depth to recover 90\% of the [CII] light is $L_{IR,min} = 1.5\times 10^{10}$ L$_{\odot}$, corresponding to a [CII] line flux of $1.4\times10^{-20}$ W m$^{-2}$. A $5\sigma$ detection of this flux demands 22 hours of integration time per beam, assuming a 1$\sigma$-1hr sensitivity of 1.4 mJy at $R = 1,000$ with dual polarization and a 12-m array composed of 50 antennas.\footnote{Sensitivities have been calculated with the ALMA Sensitivity Calculator, available online at http://almascience.eso.org/proposing/sensitivity-calculator} But, crucially for ALMA, to observe enough galaxies in each luminosity and redshift bin for this purpose requires both many tunings of the observing frequency and telescope pointings on the sky to overcome shot noise, which is the dominant source of noise in the volume of the small ALMA beam. One can estimate the fractional uncertainty on $\bar{S}_{\textrm{[CII]}}$ due to variance $\sigma_{shot}^2$ from shot noise,  and thus the number of pencil beams $N_{pencils}$ required to achieve a certain fractional uncertainty from the following:
\begin{equation}
\frac{\sigma_{shot}}{\bar{S}_{\textrm{[CII]}}} = \frac{1}{N_{pencils}^{1/2}}\frac{1}{\bar{S}_{\textrm{[CII]}}}\left(\frac{P_{\textrm{[CII]},\textrm{[CII]}}^{shot}}{V_{beam}}\right)^{1/2}
\label{eq:varshot}
\end{equation}
Above, $V_{beam}$ is the volume of the pencil beam survey, 
\begin{equation}
V_{beam} = A_{pix}^{\mathrm{ALMA}} \times \Delta r_{los}^{survey}
\end{equation}
where the physical pixel area is $A_{pix}^{\mathrm{ALMA}} = 0.0073$ (Mpc/h)$^3$, and the comoving line-of-sight distance corresponding to the frequency range of the survey is given by $\Delta r_{los}^{survey}$. As a concrete example for ALMA, for 16 GHz of backend bandwidth, translating to a redshift depth of 0.04 centered at $z = 1.2$, $V_{beam} = 0.45$ (Mpc/h)$^3$. $P_{\textrm{[CII]},\textrm{[CII]}}^{shot}$ is the shot noise as calculated from the combined B11-Spinoglio model. From this expression, we find that $N_{pencils} = 48,000$ in order to achieve $\frac{\sigma_{shot}}{\bar{S}_{\textrm{[CII]}}}$ of 10\%, which, at $1.1\times10^6$ hours of total observing time excluding overheads, would then match the fractional uncertainty on $\bar{S}_{\textrm{[CII]}}$ attained by the fiducial intensity mapping balloon experiment in 450 hours. 

We note that in the future, CCAT will be more powerful than ALMA for this experiment. While this waveband is not baselined in the first-generation spectrometer concept X-Spec,  a multi-object wideband spectrometer on CCAT will be somewhat faster than ALMA. Each CCAT backend beam is a factor of 20 less sensitive than ALMA at these frequencies (850 GHz, ALMA Band 10), but the large bandwidth eliminates the need for multiple tunings ($\sim6$ to cover the full 850 GHz band) and increases the volume of the survey.   With 100 backend beams as is baselined for an early generation X-Spec, CCAT/X-Spec has an advantage of a factor of 600 in time, more than overcoming the ALMA sensitivity advantage.   With the bandwidth and beam size included, the volume of a CCAT pencil beam is 1.73 $\times$ larger than and ALMA beam, so the number of independent beams required to overcome shot noise is smaller by this factor.

Follow-up of known continuum sources with ALMA and CCAT is a possibility to lower the time cost of blind surveys, but this then becomes a biased estimate of the mean intensity, unlike the complete measurement provided by intensity mapping experiments. A benefit of the galaxy surveys, however, is their ability to independently measure the galaxy bias on large scales by comparison to the expected dark matter power spectrum, provided that the surveys can overcome cosmic variance. One appealing scenario is, therefore, to exploit the complementarity of the different approaches and perform galaxy surveys and intensity mapping experiments in conjunction with one another. 
 
\begin {table}[t]
\begin{center}
\caption {Conversions between $t_{obs}^{vox}$ and $f_{err}$ at $z = 1.48$} \label{tab:ferrconvert} 
\begin{tabular}{ l c c c}
\hline \hline
$t_{obs}^{vox} f_{err}^2$ ($\times 10^6$) & B11 & $\alpha = -1.0$ & $\alpha$ = -2.0 \\
\hline
Atmospheric Balloon & 1.54 & 1.57 & 2.83 \\
Cryogenic Satellite & 0.0286 & 0.0292 & 0.00526 \\
\hline
 \end{tabular}
 \end{center}
 \end{table}
 
 \subsection{Probes of the power spectrum}

There may be applications---such as measuring the BAO peak or searching for primordial non-Gaussianity in large-scale structure---for which the mean intensity is not required, and the shape of the power spectrum, rather than its absolute value, is of interest. For this application, we compare the SNR on a linear-term $k$ bin (up to $k < 0.3$ h/Mpc) for both galaxy detection and intensity mapping surveys (denoted, respectively, by the subscripts ``GS" and ``IM"), with the expressions:

\begin{align}
\mathrm{SNR}_{GS} & = \frac{\sqrt{N_{modes}} }{1 + 1/(\bar{b}_i^2P_{\delta,\delta}\bar{n}_{gal} )}
\label{eq:snr_gs}
 \\
\mathrm{SNR}_{IM} & =  \frac{\sqrt{N_{modes}}}{1 + P_N/\left(\bar{S}_{i}^2\bar{b}_i^2P_{\delta,\delta}\right)} 
 \label{eq:snr_im} 
 \\
 & = \frac{\sqrt{N_{modes}}}{1 + (f_{err}^2 V_{vox}) / (\bar{b}_i^2P_{\delta,\delta})} \nonumber
\end{align}
\begin{figure}[t]
 \centering
 \includegraphics[width=0.5\textwidth]{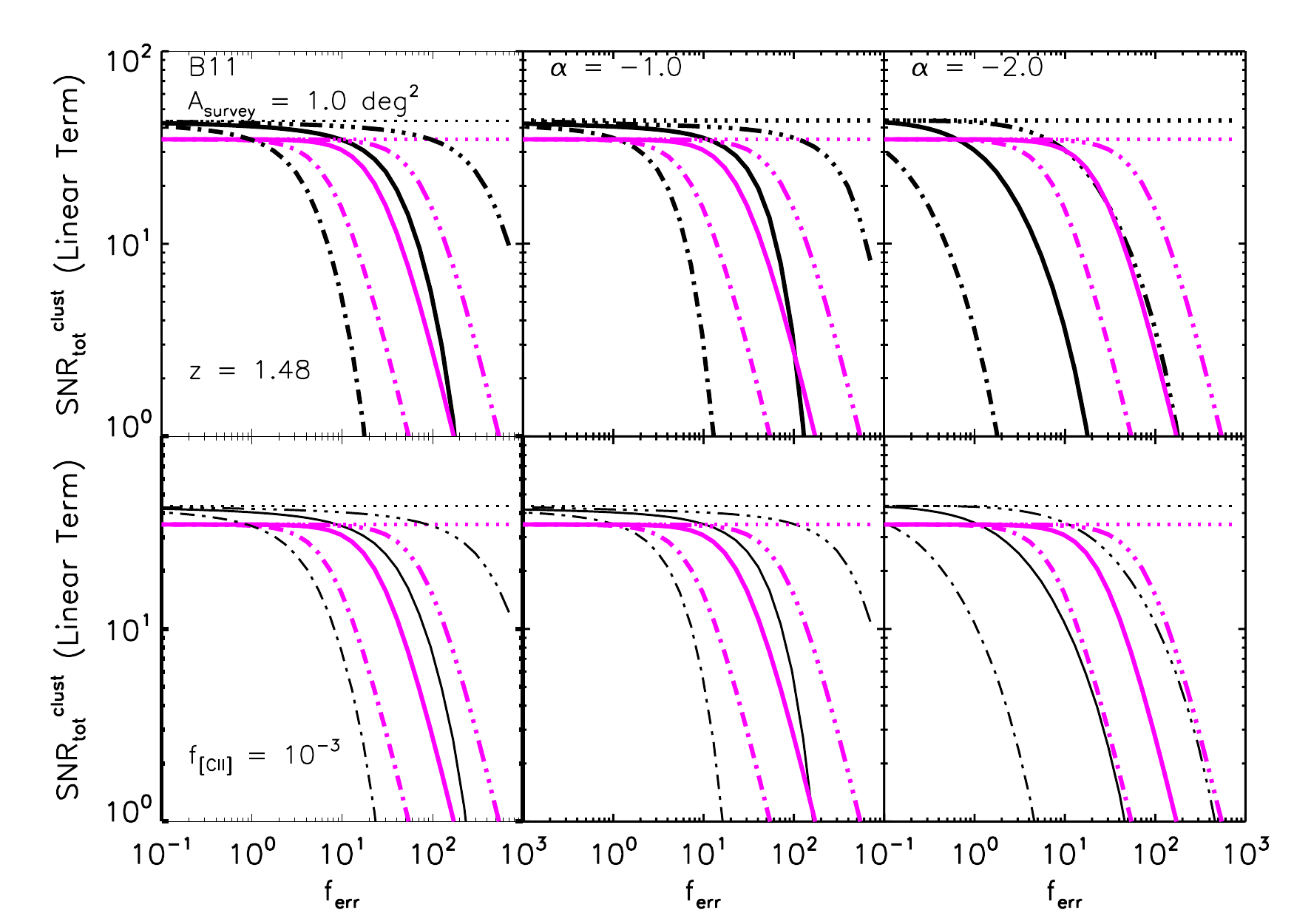}
\caption{Total signal-to-noise ratio on the linear portion of the clustering power spectrum of [CII] at $z=1.48$ as a function of the fractional error. Results are plotted for the B11 (left panels) model as well as the toy Schechter functions (middle and right panels). SNR$_{IM}$ and SNR$_{GS}$ are plotted as the magenta and black curves, respectively. Solid curves correspond to the fiducial aperture, $d_{ap}$ = 2.5 m. Dashed curves correspond to apertures scaled by a factor $\sqrt{\epsilon}$, where $\epsilon=10$ (triple-dot-dashed) and $\epsilon=0.1$. The horizontal dotted line is the maximum SNR possible for each approach as set by the number of modes in the survey volume, which is lower for the intensity mapping experiment due to our described mode removal. Results are shown for predictions of [CII] intensity based on the Spinoglio fits (top panel) and a constant ratio of $\frac{L_{\textrm{[CII]}}}{L_{IR}} = 10^{-3}$ (bottom panel).}
\label{fig:snr_vs_ferr}
\end{figure}
Equations~\ref{eq:snr_gs} and~\ref{eq:snr_im} assume that the sources in the galaxy survey have the same clustering, and thus the same $\bar{b}_i$, as the sources in the intensity mapping experiment. The quantity $\bar{n}_{gal}^{-1}$ in the expression for SNR$_{GS}$ denotes the shot noise for the galaxy survey, as $\bar{n}_{gal}$ refers to the mean number density of galaxies detected in the survey volume. 

\begin{figure}[H]
\centering
\includegraphics[width=0.5\textwidth]{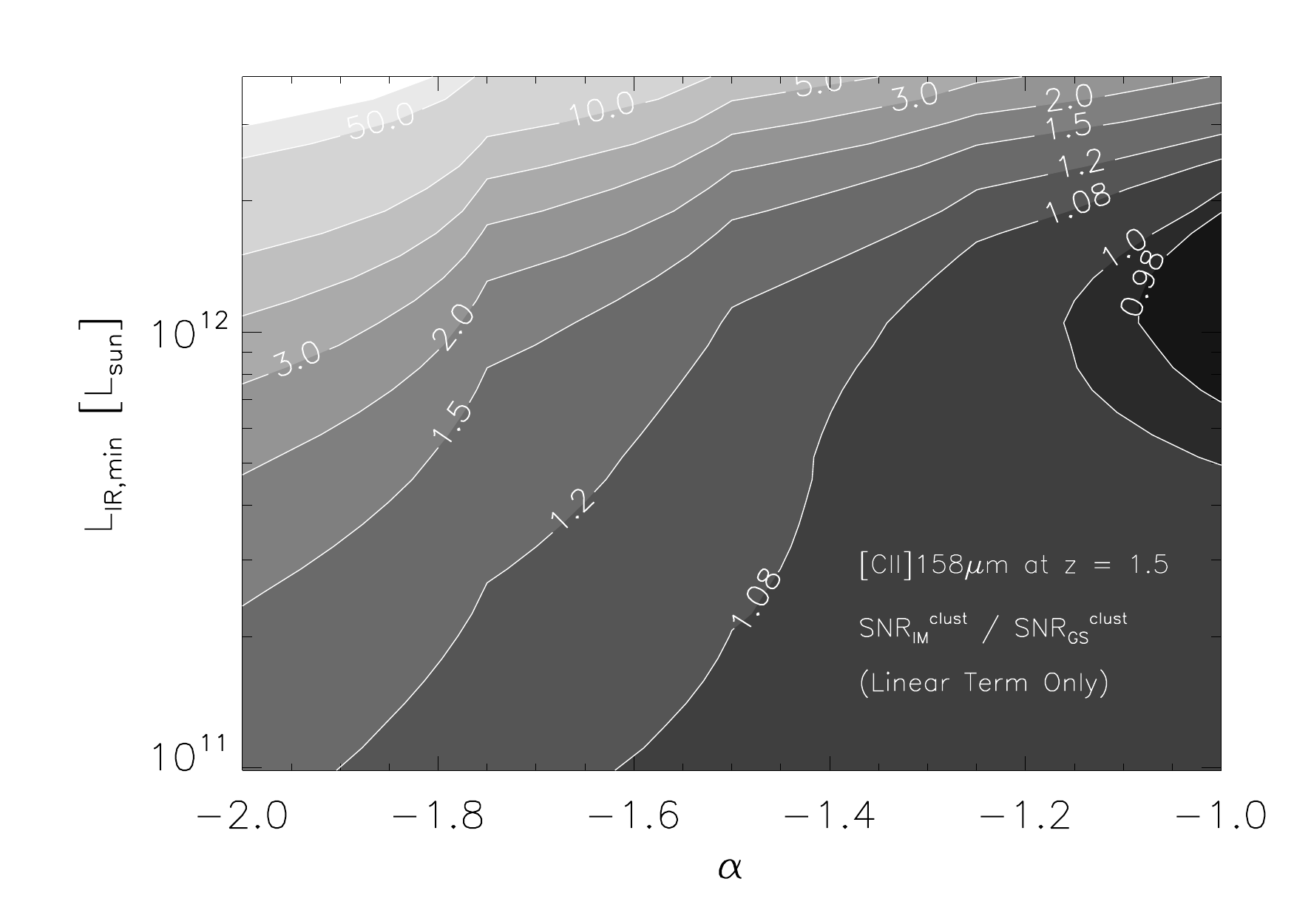}
\caption{Contours of SNR$_{IM}$/SNR$_{GS}$ for the linear term in the [CII] clustering power spectrum at $z = 1.48$, determined for a given depth (in $L_{IR}$) and IR LF faint-end slope $\alpha$.}
\label{fig:contours}
\end{figure}

Even in this limited comparison of relative SNRs, the intensity mapping often outperforms galaxy surveys, as shown in Figure~\ref{fig:snr_vs_ferr}. For the steepest faint-end slope ($\alpha = -2.0$) we have tested, SNR$_{IM} >$ SNR$_{GS}$ for all $f_{err}$ and beam sizes (i.e., telescope apertures). For the flatter LFs, there are ranges of $f_{err}$ where SNR$_{IM} >$ SNR$_{GS}$ for the fiducial case, corresponding to when the galaxy surveys are shot-noise dominated. Figure~\ref{fig:contours} summarizes the results in Figure~\ref{fig:snr_vs_ferr} by plotting contours of constant $\frac{\mathrm{SNR}_{IM}}{\mathrm{SNR}_{GS}}$ in the $L_{IR,min}-\alpha$ plane. We see in this figure that there is only a small region---occupied by very flat luminosity functions with slope $\alpha <-1.2$---where the galaxy survey measures the clustering power spectrum with greater SNR than the intensity mapping experiment. It is important to remember that while surveys may detect a large number of galaxies, and thus attain appreciable SNR$_{GS}$ on the power spectrum, the sample of detected galaxies may not yield a measurement of mean intensity, for which a large fraction of the total [CII] light must be observed (cf. Figure~\ref{fig:ngal_frac}.) 

We have focused on calculating the SNR of the linear clustering term, which constrains the total [CII] emission and the luminosity-weighted bias of the emitting galaxies. Measurements at smaller scales may help to constrain the spatial distribution of the galaxies within their host dark matter halos, by measuring the shape of the 1-halo term. 

\section{Summary and Outlook}

We have demonstrated the utility of the intensity mapping technique in measuring 3-D power spectrum of far-IR line emission at moderate redshifts, focusing on the important star-formation indicator [CII]. Fluctuations of far-IR fine-structure line intensities have been modeled by combining with the theorized dark matter power spectrum the empirically-constrained estimates of the IR luminosity from the B11 IR luminosity function and \citet{spinoglio12} line-to-$L_{IR}$ relations. We have presented predictions for the measurement of the [CII] auto-power spectrum between $0.63 < z < 1.48$, and found the power spectrum to be detectable in both clustering and shot noise terms in this redshift range with a modest, balloon-borne experimental platform, and exceptionally so with a more ambitious space-borne experimental platform. On large scales, the fact that the clustering amplitude of [CII] fluctuations is proportional to the mean [CII] intensity indicates the potential for measuring cosmic evolution of aggregate [CII]---or of any target line---emission with the line intensity mapping approach, modulo uncertainties in the bias, which may be removed by independent measures such as redshift space distortions. For the fiducial experiments considered in this paper, we have found that it would be possible to measure the [CII] luminosity density with fractional uncertainties on the order of 10\% or less. In examining the effect of luminosity function shape, telescope aperture, and fractional error (or instrument noise level) on the relative performances of intensity mapping to galaxy surveys, we have further demonstrated that, in the case where experiments with low fractional errors are not feasible, intensity mapping experiments often outperform galaxy redshift surveys when measuring the mean [CII] intensity. For steep luminosity functions, intensity mapping appears to be the only means of measuring average intensity and thus constraining the bulk of the luminosity function, as well as the optimal method of measuring the clustering power spectrum. 

Although beyond the scope of this paper, our findings here reinforce the notion that the $z > 6$ Universe presents an ideal landscape to learn about galaxy populations via intensity mapping. Strong evidence for steep ($\alpha \sim -2.0$) luminosity functions in the rest frame UV at $z\sim7$ \citep{bouwens14}, and larger voxels for a given aperture at higher redshifts, combine to position intensity mapping more favorably compared to galaxy surveys in probing the nature and clustering of the reionizing population. 

Looking to the future, the unprecedented sensitivity of background-limited spectrometer technology aboard space-borne experiments as described in this paper may become novel and important platforms to conduct large ($\sim 1,000$ deg$^2$) blind spatio-spectral surveys of far-IR line emission, and warrants further study.
\\\\
The authors thank Olivier Dor\'{e} for useful discussions, and Yan Gong for valuable comments that improved this manuscript. BU acknowledges support from the NASA GSRP Fellowship. 

\appendix

To explore the effect of the luminosity function shape on the relative performances of intensity mapping and galaxy surveys in observing the [CII] power spectrum and mean intensity of [CII] emitters, we have introduced toy models to represent different $\Phi(L_{IR}, z) \equiv \frac{\textrm{d}N}{\textrm{d}L_{IR}\textrm{d}V}$.

We parametrize our luminosity function as a Schechter function
\begin{equation}
\Phi(L_{IR},z) \textrm{d}L_{IR} = \phi_* \left(\frac{L_{IR}}{L_*}\right)^{\alpha} \exp\left(-\frac{L_{IR}}{L_*}\right) \textrm{d}L_{IR}
\end{equation}
where $\phi_*$ is the normalization for number density, $L_*$ is the characteristic luminosity at the knee, and $\alpha$ is the faint-end slope, as usual. 

Power-law luminosity functions are notoriously ill-behaved if the lower limit of integration for either the luminosity functions or its moments is extended to zero. Rather than implement a break in the power law, we simply cut it off at some $L_{IR, min}$ and choose to fix in our analysis the total IR luminosity density from galaxies as predicted by B11, denoted as $\rho_{IR}^{\textrm{B11}}$, such that
\begin{equation}
\int \textrm{d}L_{IR} \phi_* L_* \left(\frac{L_{IR}}{L_*}\right)^{\alpha+1} \exp\left(-\frac{L_{IR}}{L_*}\right) \equiv \rho_{IR}^{\textrm{B11}}
\label{eq:schechterlum}
\end{equation}
This is motivated by the observation that in many cases we do have constraints on the integrated light (from, for example, the cosmic infrared background or from the cosmic star formation rate density or the requirement of critical reionization), whereas we may not in general have detailed constraints on the distribution of light among galaxies, i.e., the shape of luminosity function.
 
The number density of sources, $n_{gal}$, can, in turn, be computed from 
\begin{equation}
n_{gal} = \int \textrm{d}L_{IR} \phi_* \left(\frac{L_{IR}}{L_*}\right)^{\alpha}\exp\left(-\frac{L_{IR}}{L_*}\right)
\end{equation}

Finally, equation~\ref{eq:schechterlum} allows us to calculate the [CII] luminosity density for each IR-normalized toy model as

\begin{equation}
\rho_{\textrm{[CII]}} = \int \textrm{d}L_{IR} \phi_* L_* \left(\frac{L_{IR}}{L_*}\right)^{\alpha+1} f_{\textrm{[CII]}} \exp\left(-\frac{L_{IR}}{L_*}\right) 
\end{equation}
where $f_{\textrm{[CII]}}$ is the fraction of IR luminosity emitted in [CII], or $\frac{L_{\textrm{[CII]}}(L_{IR})}{L_{IR}}$, described by the Spinoglio relations. Because $L_{\textrm{[CII]}}$ is slightly sublinear in $L_{IR}$, it follows that the toy models with steep faint-end slopes will produce more [CII] emission than their flatter counterparts.

\end{document}